\documentclass[
 reprint,
nofootinbib,
 amsmath,amssymb,
 aps,
]{revtex4-2}

\usepackage{graphicx, color}% Include figure files
\usepackage{dcolumn}% Align table columns on decimal point
\usepackage{bm}% bold math
\usepackage[mathlines]{lineno}% Enable numbering of text and display math
\usepackage{amsmath}
\usepackage{amssymb}
\usepackage{bm}
\usepackage{mathrsfs}
\usepackage{float}
\usepackage{soul}
\usepackage{mathrsfs}
\usepackage{lipsum}
\usepackage{hyperref}
\usepackage[normalem]{ulem}
\hypersetup{
    colorlinks=true,       % false: boxed links; true: colored links
    linkcolor=green2,          % color of internal links
    citecolor=green2,        % color of links to bibliography
    filecolor=magenta,      % color of file links
    urlcolor=green2           % color of external links
}
\usepackage{cleveref}
\usepackage{color}

\newcommand{\beq}{\begin{align}}
\newcommand{\eeq}{\end\begin{align}}

\def\lap{\lower.5ex\hbox{$\; \buildrel < \over \sim \;$}}
\def\gap{\lower.5ex\hbox{$\; \buildrel > \over \sim \;$}}

\def\be{\begin{align}}
\def\ee{\end\begin{align}}
\def\ba{\begin{eqnarray}}
\def\ea{\end{eqnarray}}

\def\b{\boldsymbol}
\def\bk{\b k}

\def\bx{\b x}

\def\by{\b y}

\def\bw{\b w}

\newcommand{\fref}[1]{Fig.~\ref{#1}}

\definecolor{rp}{cmyk}{0.2, 1, 0.6, 0}
\definecolor{rp}{cmyk}{0.2, 1, 0.6, 0}
\definecolor{green2}{cmyk}{0.27, 0, 1, 0.52}

\newcommand{\bPsi}{{\bm{\mathsf{\Psi}}}}
\newcommand{\bepsilon}{{\bm{\mathsf{\epsilon}}}}

\newcommand{\mpl}{m_{\mathrm{pl}}}
\newcommand{\schr}{\rm Schr{\"o}dinger }
\DeclareMathOperator{\Tr}{Tr}

\begin{document}

\preprint{APS/123-QED}

\title{{\textsf{\textbf{i-SPin 2}}}: An integrator for general spin-$s$
Gross-Pitaevskii systems}

\author{Mudit Jain}
\email{mudit.jain@rice.edu}
\author{Mustafa A. Amin}
\email{mustafa.a.amin@rice.edu}
\author{Han Pu}
\email{hpu@rice.edu}

\affiliation{Department of Physics and Astronomy, Rice University, Houston, Texas 77005, U.S.A.}

\date{\today}

\begin{abstract}
We provide an algorithm for evolving general spin-$s$ Gross-Pitaevskii / non-linear Schr{\"o}dinger systems carrying a variety of interactions, where the $2s+1$ components of the `spinor' field represent the different spin-multiplicity states. We consider many nonrelativistic interactions up to quartic order in the Schr{\"o}dinger field (both short and long-range, and spin-dependent and spin-independent interactions), including explicit spin-orbit couplings. The algorithm allows for spatially varying external and/or self-generated vector potentials that couple to the spin density of the field. Our work can be used for scenarios ranging from laboratory systems such as spinor Bose-Einstein condensates (BECs), to cosmological/astrophysical systems such as self-interacting bosonic dark matter. As examples, we provide results for two different setups of spin-$1$ BECs that employ a varying magnetic field and spin-orbit coupling, respectively, and also collisions of spin-$1$ solitons in dark matter. Our symplectic algorithm is second-order accurate in time, and is extensible to the known higher-order accurate methods.
\end{abstract}

\maketitle

\tableofcontents
%%%%%%%%%%%%%%%%%%%%%%%%%%%%%%%%%%%%%%%%%%%%%%%%%%%%%%%%%%

%------------------
\section{Introduction}
\label{sec:intro}
%------------------

Physical systems described by Gross–Pitaevskii equation (GPE)/non-linear \schr equation (NLSE) are ubiquitous in many areas of physics, ranging from laboratory systems such as ultracold atomic Bose-Einstein condensates (BECs)~\cite{Dalfovo:1999zz}, non-linear optics~\cite{10.1007/3-540-46629-0_9,menyuk1987nonlinear, christodoulides1988vector, rand2007observation, sun2009bound, baronio2012solutions,1973ZhETF..65..505M,1073308}, water waves~\cite{https://doi.org/10.1002/sapm1967461133,https://doi.org/10.1002/sapm1976553231,doi:10.1146/annurev.fluid.40.111406.102203}, etc., to cosmological scenarios concerning the phenomenology of cold dark matter~\cite{Turner:1983he,Press:1989id,Sin:1992bg,Goodman:2000tg,Guzman:2003kt,Amendola:2005ad,Calabrese:2016hmp,Niemeyer:2019aqm,Ferreira:2020fam,Adshead:2021kvl,Jain:2021pnk,Amin:2022pzv,Gorghetto:2022sue,Jain:2022kwq}. 

In the case of BECs in laboratory, atoms are cooled and trapped using magnetic or optical traps. With magnetic traps the various hyperfine levels of the atoms are lost and the system can be described using one component (scalar) GPE/NLSE. The use of optical traps, however, gives leverage over the different possible hyperfine levels, resulting in the so called spinor BECs \cite{PhysRevLett.80.2027,PhysRevLett.81.5257,RevModPhys.85.1191}. Such a system can be described by a multicomponent GPE/NLSE.\footnote{Throughout this work, we shall generically refer to the hyperfine state, in the context of AMO (Atomic, molecular, and optical physics) systems, as spin.} 

Depending upon the atomic species and the experimental setup, the different spin components can have many types of both short-ranged and long-ranged self-interactions (in addition to interactions with the external trapping potential and magnetic field). For instance, the long-ranged interaction could be mediated by the dipolar ($\sim 1/r^{3}$) interaction potential generated due to the spin density of the \schr field. The short-ranged self-interaction can be both spin-independent and spin-dependent.
The former is the density-density interaction of type $\sim \rho^2$ where $\rho$ is the total number density of the multicomponent \schr field. The latter can come in different varieties. One such spin-dependent interaction is the usual spin-spin interaction of type $\sim\bm{\mathcal{S}}\cdot\bm{\mathcal{S}}$, where $\bm{\mathcal{S}}$ is the intrinsic spin density of the field. Another spin-dependent interaction is the spin-singlet interaction which characterizes collisions between two particle spin singlet states. Besides such self-interactions, there are other possible interactions such as the spin-orbit interaction.

So far, various higher spin condensates have been achieved in laboratory experiments. For instance, see ~\cite{Myatt:1997zz,Lin_nature} for spin-$1/2$, see ~\cite{PhysRevLett.80.2027,Stenger_nature,PhysRevLett.87.010404} for spin-$1$, see ~\cite{PhysRevA.61.033607,PhysRevA.63.012710,PhysRevLett.92.140403,Widera_2006} for spin-$2$, and ~\cite{PhysRevA.81.042716,PhysRevLett.106.255303} for spin-$3$ condensates. Owing to their spin (hyperfine) structure, such BEC systems are promising for interesting effects such as topological spin textures~\cite{PhysRevLett.122.095301,Hall}, quantum spin hall effect and topological insulators~\cite{doi:10.1126/science.1105514,doi:10.1126/science.1148047,PhysRevLett.95.146802,doi:10.1126/science.1133734,Hsieh_nature}, atomic lasers~\cite{PhysRevLett.82.3008,Bolpasi:2013wia}, etc. See Refs.~\cite{CiCP-24-899,KAWAGUCHI2012253,RevModPhys.85.1191,Ueda_2014} and references therein for detailed reviews. Understanding the behavior of such higher spin systems from an analytical and computational standpoint is therefore highly desired. 

In the cosmological scenario, the GPE/NLSE is used to describe the cold dark matter field, and can contain both the density-density and spin-spin interactions (in the case of higher spin dark matter), besides the usual gravitational interactions. For instance in the case of vector dark matter, both of these self-interactions are present in the effective low energy regime (Higgs phase) of the Abelian Higgs model~\cite{Zhang:2021xxa,Jain:2022kwq}. Even in the case of massive spin-$2$/bi-gravity constructions~\cite{deRham:2010kj,Hassan:2011hr,Hassan:2011tf,Hassan:2011zd,Hinterbichler:2011tt,deRham:2014zqa,Schmidt-May:2015vnx}, there are quartic self-interactions of the massive spin-$2$ degree of freedom~\cite{Babichev:2016bxi}, and can very well result in spin-spin interactions in the nonrelativistic low energy effective theory (besides the density-density interactions). Our work here therefore, can naturally find its relevance in many cosmological/astrophysical scenarios.\\

In this paper, we present for the first time, a 3D numerical algorithm involving split Fourier technique, to evolve (a) \textit{arbitrary spin-$s$} condensates containing both short and long-ranged quartic self-interactions, in addition admitting (b) space and time-dependent external vector fields $\bar{\bm B}({\bm x},t)$ (but separable such that $\bar{\bm B}({\bm x},t) = {\bm B}({\bm x})f(t)$), leading to not only spatially and time varying Zeeman effects, but also (and perhaps more importantly) giving rise to spin-orbit (SO) coupling. Lastly, (c) an explicit spin-orbit coupling term that couples the spin and the center-of-mass momentum. The SO coupling, arising in setups involving multiple lasers~\cite{Lin_nature}, and its effects have been gaining much interest recently \cite{Goldman_2014,Zhai_2015}.

This work can be contrasted with the existing literature, in which some work on this front already exists: In Ref.~\cite{PhysRevE.93.053309}, similar split Fourier technique was employed for spin-$1$ GPE containing both spin-spin and density-density short-ranged self-interactions, along with a spatially uniform linear and quadratic Zeeman term. In Ref.~\cite{PhysRevE.95.013311}, the same situation was explored for a spin-$2$ system, with the addition of the spin singlet interaction term. More recently, in Ref.~\cite{SMITH2022108314} the authors presented a GPU-assisted approach to
accelerate solving 2-D spin-$1/2$ GPE/NLSE. Our work in this paper differs from the existing literature in the 3 points listed in the previous paragraph. Also, contrary to a previous work by some of us~\cite{Jain:2022agt} where the $n$-component \schr field had an SO($n$) symmetry, systems of consideration in this paper are describable by a $2s+1$ component \schr field, with components characterizing the different spin multiplicity states.\\

Our symplectic (Unitary) algorithm employs the split-Fourier step technique in which the field evolution over a time step is broken into a half `drift' piece, followed by a `kick' piece, and then another half `drift' piece. In the drift pieces, the field is evolved using the drift Hamiltonian density that contains the usual Laplacian term together with the SO coupling term. In the kick piece, the field is evolved using the interaction Hamiltonian density which contains all of the rest of the interaction terms. By explicitly constructing the Unitary  evolution matrices in both the drift and the kick steps, we present a symplectic time-reversible algorithm. The accuracy of the field evolution in this algorithm is $\mathcal{O}(\epsilon^2)$ (where $\epsilon$ is the time discretization step), which can be upgraded towards higher order symplectic integrators that employ the split-step technique.\\

The paper is organized as follows. In section~\ref{Sec:model} we begin by laying out the general (non-relativistic) spin-$s$ Schr\"{o}dinger system containing all interactions of interest, including interactions with external scalar and vector fields. Then in section~\ref{sec:evolution} we work out the analytical solution for the field evolution due to both the drift and kick Hamiltonian densities, with the most non-trivial bit being the exponential of spin matrices. In~\ref{sec:matrix_exp} we present the general scheme of exponentiating arbitrary spin-$s$ matrices, and provide explicit results for spin-$1$, $2$, and $3$ case in appendix~\ref{app_spinmatrices_AMO}. With the analytical solution at hand, the general algorithm scheme is provided in section~\ref{sec:algo_summary}. Our work has a broad domain of applicability, ranging from AMO physics in the laboratory to self interacting fuzzy dark matter in cosmology. We discuss this in section~\ref{sec:scope}, and present some simulation results for three example scenarios, demonstrating the effects of some of the interactions of interest. Section~\ref{Sec:summary} presents the summary of our work. In Appendix~\ref{app_spinmatrices_FT} we present the conventional forms for the spin matrices for spin-$1$ and spin-$2$ cases that are more suited for cosmology.\\

\noindent{\bf{Units and conventions:}} Throughout the paper and unless explicitly written, we work in natural units where $\hbar = 1 = c$. We also assume Einstein summation convention.

%--------------------%
\section{Spin-$s$ Gross-Pitaevskii / Schr\"{o}dinger system}
\label{Sec:model}
%--------------------%

\subsection{Action and equation of motion}

Our system comprises of a $2s+1$-component Schr\"{o}dinger field $\Psi = (\psi_{s},\psi_{s-1},...,\psi_{-s})$ of mass $\mu$, where different components represent the various spin multiplicity levels. More formally, $\Psi$ transforms as a vector / `spinor', in the $2s+1$ dimensional irreducible Unitary representation of the SO($3$) group.\footnote{The quotes on `spinor' is to highlight that it is not the fermion spinor that is usually referred to in the context of particle physics/quantum field theory. We will drop the quotes in the rest of the paper.} For this system, we consider the following general action up to quartic order in the field $\Psi$, including all the relevant self-interactions (to leading order in the nonrelativistic limit)
\begin{align}\label{eq:nonrel_action}
    \mathcal{S}_{\rm nr} =& \int \mathrm dt\,\mathrm d^3x\Biggl[\frac{i}{2}\psi^{\dagger}_n\dot{\psi}_n + \mathrm{c.c.} - \frac{1}{2\mu} \nabla\psi^{\dagger}_{n}\cdot\nabla\psi_{n}\nonumber\\
    &\qquad\qquad - \mu\rho V(\bm x) - \gamma\,\bm{\mathcal{S}}\cdot\bar{\bm B}(\bm x, t) - V_{\rm nrel}(\rho,\bm{\mathcal{S}})\nonumber\\
    &\qquad\qquad - \frac{\xi}{2}\frac{1}{(2s+1)}|\psi_n\,\hat{A}_{nn'}\psi_{n'}|^2\nonumber\\
    &\qquad\qquad + i\,g_{ij}\,\psi^{\dagger}_{n}\,[\hat{S}_i]_{nn'}\,\nabla_j\,\psi_n \Biggr]\,,
\end{align}
with $\bar{\bm B}(\bm x, t) = f(t){\bm B}(\bm x)$, and
\begin{align}\label{eq:Vnr_gen}
   \qquad V_{\rm nrel}(\rho,\bm{\mathcal{S}}) = -\frac{1}{2\mu^2}\left[\lambda\rho^2 + \alpha\,(\bm{\mathcal{S}}\cdot\bm{\mathcal{S}})\right]\,.
\end{align}
The first two terms in the action~\eqref{eq:nonrel_action} dictate the usual free field evolution (of each of the field component $\psi_{m}$ where $m \in [-s,s]$). The third and fourth terms account for interactions of the field with the external scalar trapping potential $V(\bm x)$ and vector field $\bar{\bm B}(\bm x, t)$, coupling to the number density $\rho = \psi_{n}^\dagger\psi_{n}$ and spin density $\bm{\mathcal{S}} = \psi^{\ast}_{n}\,\bm{\hat{S}}_{nn'}\,\psi_{n'}$ respectively.\footnote{Apart from the linear Zeeman term $\sim \psi^{\dagger}\,{\bm B}\cdot\bm{S}\,\psi$, there could also be a quadratic Zeeman term $\sim \psi^{\dagger}\,({\bm B}'\cdot\hat{\bm S})^2\,\psi$ which we do not consider explicitly. While it is trivial to include if ${\bm B}'$ is homogeneous, for non-homogeneous case it should also not be difficult to include, using the general spin matrix exponential scheme presented in section~\ref{sec:matrix_exp} (applied towards exponentiation of the square of spin matrices).} Here $\hat{S}_x = \hat{x}\cdot\bm{\hat{S}}$ (and similarly for $y$ and $z$), are the $(2s+1)\times(2s+1)$ dimensional spin matrices with the usual commutation relations
\begin{align}
    [\hat{S}_x,\hat{S}_y] = i\hat{S}_z \quad {\rm with\;all\;cyclic\;permutations}\,.
\end{align}
The fifth term in the action~\eqref{eq:nonrel_action} accounts for quartic self-interactions of the {\schr}field, that can depend on both number density and spin density (as seen in~\eqref{eq:Vnr_gen} explicitly). The sixth term in Eq.~\eqref{eq:nonrel_action} accounts for the $2$ body spin singlet interaction, where the total spin multiplicity due to both the incoming and outgoing states add to zero. The spin singlet matrix is real and has the following properties
\begin{align}\label{eq:A_props}
    \hat{A}^{-1} = \hat{A} = \hat{A}^{T} \;;\; \hat{A}\,\hat{S}_i\,\hat{A} = -\hat{S}^{\ast}_i \;;\; \psi^{T}\hat{A}\,\hat{S}_i\psi = 0\,,
\end{align}
and an explicit form for it is given ahead in Eq.~\eqref{eq:spin.singlet.matrix}. Finally, the last (seventh) term in Eq.~\eqref{eq:nonrel_action} accounts for SO coupling where $g_{ij}$ are real constants, with $i$ and $j$ being spatial indices. Specifically, $g_{ij} \propto \epsilon_{ij3}$ (the Levi Civita symbol) gives the well known Rashba SO coupling, usually studied in the context of two-component BECs. Similarly, $g_{ij} \propto |\epsilon_{ij3}|$ gives the Dresselhaus SO coupling. The action~\eqref{eq:nonrel_action} leads to the following equation of motion, the Schr\"{o}dinger/Gross-Pitaevskii equation:
\begin{align}\label{eq:masterS1}
    i\partial_t\psi_{n} &= \Biggl[\delta_{nn'}\left(-\frac{1}{2\mu}\nabla^2\right) + \left(\mu\,V(\bm x) -  \frac{\lambda}{\mu^2}\,\rho\right)\delta_{nn'}\nonumber\\
    &\qquad + \gamma\,f(t){\bm B}(\bm x)\cdot\bm{\hat{S}}_{nn'}\;-\; \frac{\alpha}{\mu^2}\,\bm{\mathcal{S}}\cdot\bm{\hat{S}}_{nn'} \nonumber\\
    &\qquad + \frac{\xi}{2s+1}\hat{A}_{nm}\,\psi^{\ast}_m\psi_{m'}\,\hat{A}_{m'n'}\nonumber\\
    &\qquad - i\,g_{ij}\,[\hat{S}_i]_{nn'}\,\nabla_j\Biggr]\psi_{n'}\,.
\end{align}
We break the Hamiltonian density (the term in the bracket above) into a drift and a kick piece as follows: 
\begin{align}\label{eq:H_drift_n_kick}
    [\mathcal{H}_{\rm drift}]_{nn'} &\equiv \delta_{nn'}\left(-\frac{1}{2\mu}\nabla^2\right) - i\,g_{ij}\,[\hat{S}_i]_{nn'}\,\nabla_j\,,\nonumber\\
    %%%%%%%
    [\mathcal{H}_{\rm kick}]_{nn'} &\equiv \left(\mu\,V(\bm x) -  \frac{\lambda}{\mu^2}\,\rho\right)\delta_{nn'} + \gamma\,f(t){\bm B}(\bm x)\cdot\bm{\hat{S}}_{nn'}\nonumber\\
    &\quad - \frac{\alpha}{\mu^2}\,\bm{\mathcal{S}}\cdot\bm{\hat{S}}_{nn'} + \frac{\xi}{2s+1}\hat{A}_{nm}\,\psi^{\ast}_m\psi_{m'}\,\hat{A}_{m'n'}\,.
\end{align}
Throughout this work, we work in the $z$ basis. That is, the spin matrix $\hat{S}_z$ is diagonal, with the eigenvalues $m \in [-s,s]$ along the diagonal. Explicitly, and more suited for AMO systems, the spin matrices $\hat{S}_i$ and the spin singlet matrix $\hat{A}$ take the following conventional forms respectively
\begin{align}\label{eq:spin.matrices}
    [\hat{S}_x]_{nn'} &= \frac{1}{2}(\delta_{n, n'+1} + \delta_{n+1,n'})\sqrt{s(s+1)-nn'}\,,\nonumber\\
    [\hat{S}_y]_{nn'} &= \frac{1}{2i}(\delta_{n, n'+1} - \delta_{n+1,n'})\sqrt{s(s+1)-nn'}\,,\nonumber\\
    [\hat{S}_z]_{nn'} &= \delta_{n,n'}\,n\,,
\end{align}
\begin{align}\label{eq:spin.singlet.matrix}
    \hat{A}_{nn'} = (-1)^{s-n}\delta_{n,-n'}\,.
\end{align}
From a relativistic field theory/particle physics point of view on the other hand (more suited for cosmology), the spin matrices take different forms. See appendix~\ref{app_spinmatrices_FT} for details, and section~\ref{sec:cosmology} for a discussion of cosmological applications of our work.

\subsubsection*{Long-range self potentials}

The external potentials $V$ and $\bar{\bm B}$ can also be easily appended with self-generated ones, suitable for different applications. Explicitly for purposes in contemporary (ultra-)light dark matter cosmology, $V({\b x}) \rightarrow \Phi(t,{\b x})$ where $\Phi(t,{\b x})$ is the Newtonian potential, given by the Poisson equation 
\begin{align}\label{eq:gravity_Poissoneqn}
    \nabla^2\Phi(t,{\bm x}) = 4\pi\mu G\rho(t,\bm x)\,.
\end{align}
Similarly in the context of AMO systems where atomic dipolar interactions are present, $\bar{\bm B}(t,{\b x}) \rightarrow \nabla a(t,{\b x})$ where $a$ is a scalar field obeying the following Poisson equation
\begin{align}\label{eq:dipolarfield_Poissoneqn}
    \nabla^2 a(t,{\bm x}) = \gamma\,\nabla\cdot\bm{\mathcal{S}}(t,{\bm x})\,.
\end{align}

\subsection{Conserved quantities and continuity equations}\label{sec:conserved.quant}

The only conserved quantity, associated with our non-relativistic system~\eqref{eq:nonrel_action} is
the total particle number $N$ (or equivalently the total mass $M = \mu N$)
\begin{align}
\label{eq:conserved}
    N &=\int \mathrm d^3x \,\rho\,.
\end{align}
Furthermore, the local continuity equations for the number and spin densities are
\begin{align}\label{eq:cont_eqns}
&\partial_t \rho +\nabla\cdot \bm{\mathcal{J}}_{ll} = 0\,,\qquad {\rm and}\nonumber\\
&\partial_t \bm{\mathcal{S}} + \bm{\hat{S}}_{nn'}\,(\nabla\cdot\bm{\mathcal{J}}_{n'n}) = \left(\gamma\,f\,{\bm B}\times\bm{\mathcal{S}}\right) + \bm{\mathcal{J}}'\,
\end{align}
respectively, where $\bm{\mathcal{J}}_{mn}$ is a general Schr\"{o}dinger current matrix given by
\begin{align}\label{eq:tensor_current}
\bm{\mathcal{J}}_{n'n} \equiv \frac{i}{2\mu}\left[\psi_{n'}\nabla\psi^{\ast}_{n} - \psi^{\ast}_{n}\nabla\psi_{n'}\right]\,,
\end{align}
and $\bm{\mathcal{J}}'$ is the SO current term
\begin{align}
    \mathcal{J}'_{i} = i\,g_{kj}\,\epsilon_{k\ell i}\,\nabla_{j}\mathcal{S}_{\ell}\,.
\end{align}
In the spin continuity equation, the first term on the right hand side gives rise to the well known spin precession effect, while the second term dictates the SO coupling effect. For the case when $\bar{\bm B}$ is time-independent (meaning $f = const$.), the total energy in the system is also conserved. Furthermore, if $\bar{\bm B}$ and $V$ are constants in both space and time and $g_{ij}=0$, the total linear momentum and the parallel component of the total angular momentum (parallel to $\bar{\bm B}$) are also conserved (with orbital and spin angular momentum conserved separately). Also, the magnitudes of the total orbital and spin angular momentum are (separately) conserved. Adding the self generated Newtonian potential or the dipolar potential does not change these results.

%--------------------%
\section{Evolution of the Schr{\"o}dinger field}\label{sec:evolution}
%--------------------%

We employ a split Fourier algorithm in which the evolution of the \schr field is broken into two parts: drift and kick as dictated by the respective Hamiltonian densities in~\eqref{eq:H_drift_n_kick}. 
In this section we present the 
general scheme of the \schr field evolution for arbitrary integer spin fields, due to both the drift and kick Hamiltonian densities.

\subsection{Evolution due to the drift Hamiltonian density}

This evolution, c.f.~\eqref{eq:H_drift_n_kick}, is dictated by the following differential equation
\begin{align}
    i\partial_t\psi_n = \Biggl[\delta_{nn'}\left(-\frac{1}{2\mu}\nabla^2\right) - i\,g_{ij}\,[\hat{S}_i]_{nn'}\,\nabla_j\Biggr]\psi_{n'}\,.
\end{align}
Evidently, the evolution is most easily performed in Fourier space. With $\tilde{\psi}(\bm k)$ as the Fourier transformed field (and $e^{i{\bm k}\cdot{\bm x}}$ as the forward Fourier coefficient), we have
\begin{align}
    i\partial_t\tilde{\psi}_n = \Biggl[\delta_{nn'}\left(\frac{k^2}{2\mu}\right) - \,g_{ij}\,[\hat{S}_i]_{nn'}\,k_j\Biggr]\tilde{\psi}_{n'}\,,
\end{align}
giving
\begin{align}\label{eq:drift_evolution}
    \tilde{\psi}_n(t) = e^{-i(t-t_0)k^2/2\mu}[e^{i(t-t_0)g_{ij}k_{j}\hat{S}_{i}}]_{nn'}\tilde{\psi}_{n'}(t_0)\,.
\end{align}
Note that the order of the two exponentials here does not matter since the respective operations commute. Nevertheless, the most non-trivial task in the above is the matrix exponentiation, needed for the SO coupling term. We present matrix exponentials for the general spin-$s$ case in sec.~\ref{sec:matrix_exp} ahead.

\subsection{Evolution due to the kick Hamiltonian density}

Next comes the contribution from the kick Hamiltonian density, c.f. Eq.~\eqref{eq:H_drift_n_kick}, dictating the following differential evolution
\begin{align}\label{eq:kick_diffevolve}
    i\partial_t\psi_n &= \Biggl[\left(\mu V - \frac{\lambda}{\mu^2} \rho\right)\delta_{nn'} + \gamma\,f(t)\,{\bm B}\cdot\bm{\hat{S}}_{nn'}\nonumber\\
    &\quad - \frac{\alpha}{\mu^2}\,\bm{\mathcal{S}}\cdot\bm{\hat{S}}_{nn'} + \frac{\xi}{2s+1}\hat{A}_{nm}\,\psi^{\ast}_m\psi_{m'}\,\hat{A}_{m'n'}\Biggr]\psi_{n'}\,.
\end{align}
Here we have suppressed the explicit spatial dependence of $V$ and ${\bm B}$ to be concise in our notation. To get the analytical solution to the above differential equation, we can handle the four different terms on the right hand side in steps. For this purpose, it will be useful to define the following exponential operators
\begin{align}\label{eq:B_G_define}
    \hat{\mathcal{B}}_{nn'}(t,t_0) &= [e^{-i\gamma\,F(t,t_0)\,{\bm B}\cdot\bm{\hat{S}}}]_{nn'}\nonumber\\
    \hat{\mathcal{G}}_{nn'}(t,t_0) &= [e^{i(\alpha/\mu^2)(t-t_0)\bm{\mathcal{S}}(t_0)\cdot\bm{\hat{S}}}]_{nn'}\,,
\end{align}
where for ease of notation we have defined $F(t,t_0) \equiv \int^{t}_{t_0}\mathrm{d}\tau f(\tau)$, and explicitly state the property
\begin{align}\label{eq:A_prop2}
    [e^{ih(t)\,\bm{v}\cdot\hat{\bm S}}]^{T}\,\hat{A}\,e^{ih(t)\,\bm{v}\cdot\hat{\bm S}} = \hat{A}\,
\end{align}
where ${\bm v}$ is any (time-independent) vector. The above can be seen to hold true on account of the properties~\eqref{eq:A_props}.\\

To begin with, first note that the number density is constant throughout the kick evolution~\eqref{eq:kick_diffevolve}. This can be seen directly by recalling that there are no \schr currents in the kick step (c.f. Eq.~\eqref{eq:cont_eqns} with $\bm {\mathcal{J}} = 0$). To account for the evolution due to ${\bm B}$, we plug the following ansatz 
\begin{align}
    &\psi_{n}(t) = e^{-i(t-t_0)(\mu V-(\lambda/\mu^2)\rho)}\hat{\mathcal{B}}_{nn'}(t,t_0)\phi_{n'}(t)
\end{align}
into Eq.~\eqref{eq:kick_diffevolve}, to have the remaining evolution due to the spin-spin and spin singlet self-interaction:
\begin{align}\label{eq:evolve:phi1}
    i\partial_t\phi_n &= -\frac{\alpha}{\mu^2}\hat{\mathcal{B}}^{\dagger}_{n\ell}(t,t_0)\,[\bm{\mathcal{S}}(t)\cdot\bm{\hat{S}}]_{\ell m'}\,\hat{\mathcal{B}}_{m'n'}(t,t_0)\phi_{n'}\nonumber\\
    &\quad + \frac{\xi}{2s+1}\hat{A}_{n \ell'}\,\phi^{\ast}_{\ell'}\phi_{\ell}\,\hat{A}_{\ell n'}\phi_{n'}\,.
\end{align}
Here in the second line, we made use of the properties~\eqref{eq:A_props}, in order to simplify the term $[\hat{\mathcal{B}}^{\dagger}\hat{A}\hat{\mathcal{B}}^{\ast}]_{n\ell'} = - [\hat{\mathcal{B}}^{\dagger}\hat{\mathcal{B}}\hat{A}]_{n\ell'} = -\hat{A}_{n\ell'}$, and also $\hat{\mathcal{B}}_{m\ell}[\hat{A}\hat{\mathcal{B}}]_{mn'} = - [\hat{\mathcal{B}}^{\dagger}\hat{\mathcal{B}}]_{mn'}\hat{A}_{m\ell} = -\hat{A}_{n'\ell} = -\hat{A}_{\ell n'}$.

Now, the matrix in the first term of Eq.~\eqref{eq:evolve:phi1} is nothing but the backwards evolution of the spin density $\bm{\mathcal{S}}(t)$, giving the spin density at the initial instant $\bm{\mathcal{S}}(t_0)$. To see this, let us decompose $\bm{\mathcal{S}}(t)$ and $\bm{\hat{S}}$
into a parallel and a perpendicular component, with respect to the external field ${\bm B}$. Owing to the spin precession during the kick step, dictated by the only non-zero (first) term on the right hand side of the spin density continuity equation (c.f. Eq.~\eqref{eq:cont_eqns} with $\bm{\mathcal{J}} = 0 = \bm{\mathcal{J}}'$), the parallel spin density $\mathcal{S}_{||}$ does not change. However the perpendicular components of the spin density do evolve. Decomposing these perpendicular components into raising and lowering pieces, $\mathcal{S}_{+}$ and $\mathcal{S}_{-}$ (using the usual convention of right handed orientation\footnote{At any spatial location, calling the direction of ${\bm B}$ as $x_3$, the perpendicular spin matrices $\hat{S}_{x_1}$ and $\hat{S}_{x_2}$ can be used to define raising and lowering spin matrices as $\hat{S}_{\pm} \equiv \hat{S}_{x_1} \pm i\hat{S}_{x_2}$. Consecutively, we also define $\mathcal{S}_{\pm} \equiv \mathcal{S}_{x_1} \pm i\mathcal{S}_{x_2}$.}), we get the following:
\begin{widetext}
\begin{align}\label{eq:evolve:phi2}
    \hat{\mathcal{B}}^{\dagger}_{m\ell}(t,t_0)\,[\bm{\mathcal{S}}(t)\cdot\bm{\hat{S}}]_{\ell n}\,\hat{\mathcal{B}}_{nm'}(t,t_0) = \left[ \mathcal{S}_{||}(t_0)\hat{S}_{||} + \frac{1}{2}[e^{i{B}{\hat{S}_{||}}\,F(t,t_0)}] [\mathcal{S}_{-}(t)\hat{S}_{+} + \mathcal{S}_{+}(t)\hat{S}_{-}] [e^{-i{B}{\hat{S}}_{||}\,F(t,t_0)}]\right]_{m m'}\,.
\end{align}
\end{widetext}
This can be simplified further. First note that the spin precession throughout the kick step, due to ${\bm B}$, goes as follows
\begin{align}\label{eq:iden1}
    \mathcal{S}_{+}(t) &= \mathcal{S}_{+}(t_0)\,e^{i B F(t,t_0)}\,,\nonumber\\
    \mathcal{S}_{-}(t) &= \mathcal{S}_{-}(t_0)\,e^{-i B F(t,t_0)}\,.
\end{align}
Upon using this together with the identity
\begin{align}\label{eq:iden2}
    [e^{i{B}{\hat{S}_{||}}\,F(t,t_0)}] \hat{S}_{\pm} [e^{-i{B}{\hat{S}}_{||}\,F(t,t_0)}] = \hat{S}_{\pm}\,e^{\pm i B F(t,t_0)}
\end{align}
in Eq.~\eqref{eq:evolve:phi2}, the time dependence of the spin density drops out, giving
\begin{align}\label{eq:spin_simplify}
    \hat{\mathcal{B}}^{\dagger}_{m\ell}(t,t_0)\,[\bm{\mathcal{S}}(t)\cdot\bm{\hat{S}}]_{\ell n}\,\hat{\mathcal{B}}_{nm'}(t,t_0) = [\bm{\mathcal{S}}(t_0)\cdot\bm{\hat{S}}]_{m m'}\,.
\end{align}
With this simplification, we now use the ansatz
\begin{align}
    \phi_n(t) = \hat{\mathcal{G}}_{nn'}(t,t_0)\,\chi_{n'}(t)
\end{align}
in Eq.~\eqref{eq:evolve:phi1} (appended by Eq.~\eqref{eq:spin_simplify}), to give
\begin{align}
    i\partial_t\chi_n = \frac{\xi}{2s+1}\hat{A}_{n\ell'}\chi^{\ast}_{\ell'}\chi_{\ell}\,\hat{A}_{\ell n'}\chi_{n'}\,.
\end{align}
Here once again, we have used the properties~\eqref{eq:A_props} to simplify the terms $[\hat{\mathcal{G}}^{\dagger}\hat{A}\hat{\mathcal{G}}^{\ast}]_{n\ell'} = - [\hat{\mathcal{G}}^{\dagger}\hat{\mathcal{G}}\hat{A}]_{n\ell'} = -\hat{A}_{n\ell'}$, and $\hat{\mathcal{G}}_{m\ell}[\hat{A}\,\hat{\mathcal{G}}]_{mn'} = - [\hat{\mathcal{G}}^{\dagger}\,\hat{\mathcal{G}}]_{mn'}\,\hat{A}_{m\ell} = -\hat{A}_{n'\ell} = -\hat{A}_{\ell n'}$. From the above equation for $\chi$, we can first obtain the evolution equation for the quantity $q \equiv \chi^{T}\hat{A}\chi = \psi^T\hat{A}\psi$. (The second equality holds true on account of the property~\eqref{eq:A_prop2}.) We find that it simply rotates as a phasor: $q(t) = q(t_0)e^{-2i(t-t_0)\xi\rho/(2s+1)}$. With this, we use the ansatz $\chi(t) = \eta(t)e^{-i(t-t_0)\xi\rho/(2s+1)}$ to get the following equation for $\eta$:
\begin{align}
    i\partial_t\eta_n = \frac{\xi}{2s+1}\left(q(t)\hat{A}_{n\ell'}\eta^{\ast}_{\ell'} - \rho\,\eta_{n}\right)\,.
\end{align}
This has the following solution
\begin{align}
    \eta_n(t) &= \hat{\mathcal{U}}'_{n\ell}(t,t_0)\psi_{\ell}(t_0)\,
\end{align}
where the operator $\hat{\mathcal{U}}'$ is given in Eq.~\eqref{eq:kick_solution} ahead, and we have set $\eta_{n}(t_0) = \psi_{n}(t_0)$ without loss of generality.\\

In summary, combining all of the above pieces together, the full kick evolution becomes
\begin{align}\label{eq:kick_solution}
    &\psi_{m}(t) = \hat{\mathcal{U}}_{mn}(t-t_0)\,\hat{\mathcal{U}}'_{n\ell}(t-t_0)\,\psi_{\ell}(t_0)\,,\quad {\rm where}\nonumber\\
    &\hat{\mathcal{U}}_{mn}(t-t_0) = e^{-i(t-t_0)\left(\mu V - \left(\frac{\lambda}{\mu^2} - \frac{\xi}{(2s+1)}\right)\rho\right)}\times\nonumber\\
    &\qquad\qquad\qquad\qquad \hat{\mathcal{B}}_{m\ell}(t,t_0)\,\hat{\mathcal{G}}_{\ell n}(t,t_0),\nonumber\\
    &\hat{\mathcal{U}}'_{n\ell}(t,t_0) = \Biggl[\cos\left(\frac{\xi\rho_{q}}{2s+1}(t-t_0)\right)\delta_{n\ell}\nonumber\\
    &\qquad\qquad\qquad + \frac{i}{\rho_{q}}\sin\left(\frac{\xi\rho_{q}}{2s+1}(t-t_0)\right)\times\nonumber\\
    &\qquad\qquad\qquad \Bigl(\rho\,\delta_{n\ell} - \hat{A}_{nn'}\psi^{\ast}_{n'}(t_0)\psi_{\ell'}(t_0)\,\hat{A}_{\ell'\ell}\Bigr)\Biggr]\,,
\end{align}
and where $\hat{\mathcal{B}}$ and $\hat{\mathcal{G}}$ are defined in Eq.~\eqref{eq:B_G_define}, and $\rho_{q} \equiv \sqrt{\rho^2 - |q(t_0)|^2} = \sqrt{\rho^2 - |\psi_{n}(t_0)\hat{A}_{nm}\psi_{m}(t_0)|^2}$. This is our main equation for the evolution of the field $\psi$ under the kick Hamiltonian density. It is important to note the order of the exponentials $\hat{\mathcal{B}}(t,t_0) = e^{-i\gamma\,{\bm B}\cdot\bm{\hat{S}}\,F(t,t_0)}$ and $\hat{\mathcal{G}}(t,t_0) = e^{i(\alpha/\mu^2)(t-t_0)\bm{\mathcal{S}}(t_0)\cdot\bm{\hat{S}}}$ in the above evolution equation. Unless ${\bm B}$ and $\bm{\mathcal{S}}$ are parallel, reversing the order leads to incorrect evolution since ${\bm B}\cdot\bm{\hat{S}}$ and $\bm{\mathcal{S}}\cdot\bm{\hat{S}}$ do not commute in general.\\

With the exact evolution for both the drift and the kick steps, Eq.~\eqref{eq:drift_evolution} and Eq.~\eqref{eq:kick_solution} respectively, we now require the analytical form for the matrix exponential $e^{-i\beta\,{\bm n}\cdot\bm{\hat{S}}}$ for a general spin-$s$ system. Here ${\bm \beta} = \beta {\bm n}$ could be any function of ${\bm k}$ or ${\bm x}$ (relevant for SO drift and kick terms respectively). For the SO term in the drift evolution, $\beta\,n_i = -(t-t_0)g_{ij}k_j$, while for the magnetic field coupling and spin-spin interaction in the kick evolution, we have $\beta\,n_i = \gamma F(t,t_0)B_i$ and $\beta\,n_i = -(\alpha/\mu^2)(t-t_0)\mathcal{S}_i$ respectively. We pursue the relevant exercise in the next section.

%------------------
\section{Matrix exponential for general spin-$s$}
\label{sec:matrix_exp}
%------------------

For any arbitrary spin-$s$, the matrix exponential in general must take the following form
\begin{align}\label{eq:ansatz_exp}
    e^{-i\beta\,\hat{\bm n}\cdot\bm{\hat{S}}} &= \mathbb{I} + i\sum^{s}_{\ell = 1}(\hat{\bm n}\cdot\bm{\hat{S}})^{2\ell-1}\Biggl[\sum^{s}_{m=1} a_{m\ell}\sin m\beta\Biggr]\nonumber\\
    &\qquad + \sum^{s}_{\ell = 1}(\hat{\bm n}\cdot\bm{\hat{S}})^{2\ell}\Biggl[\sum^{s}_{m=0} b_{m\ell}\cos m\beta\Biggr]\,.
\end{align}
Here $a_{m\ell}$ and $b_{m\ell}$ are real coefficients and $\mathbb{I}$ the $(2s+1)$-dimensional identity matrix. The reason that the above form must hold is three fold: (1) The conjugate transpose of the exponential must be the same as $\beta \rightarrow -\beta$; (2) all possible frequencies, $m \in [0,s]$, must appear in the expansion; and (3) the maximum power required of the matrix $(\hat{\bm n}\cdot\bm{\hat{S}})$ is $2s$, since all higher powers of this matrix can be written as linear combinations of $\mathbb{I}$, $(\hat{\bm n}\cdot\bm{\hat{S}})$, $(\hat{\bm n}\cdot\bm{\hat{S}})^2$, and so on up to $(\hat{\bm n}\cdot\bm{\hat{S}})^{2s}$ by virtue of Cayley-Hamilton theorem.\\

With the above form, the explicit values of the $s^2$ number of $a_{m\ell}$ and $s(s+1)$ number of $b_{m\ell}$ can be determined by matching the Taylor expansion of the exponential in $\beta$ (only up to $\beta^{2s}$) on the left hand side, with the similar Taylor expansion of the series form in the right hand side of Eq.~\eqref{eq:ansatz_exp}. For this matching purpose, it is easiest to work with $\hat{\bm n} = \hat{\bm z}$, since in our working $z$ basis $\hat{S}_{z}$ is diagonal and equal to the last expression in Eq.~\eqref{eq:spin.matrices}. Upon performing this matching exercise, we get
\begin{align}\label{eq:a_relations}
    &\sum^{s}_{\ell=1}n^{2(\ell-r-1)}\Biggl[\sum^{s}_{m=1}a_{m\ell}\,m^{2r+1}\Biggr] = -1 \nonumber\\
    &\forall \;\; n = \{1,...,s\} \;\; \mathrm{and} \;\; r = \{0,1,...,s-1\}
\end{align}
from the odd terms in $\beta$ (i.e. from the sine terms), while 
\begin{align}\label{eq:b_relations}
    &\sum^{s}_{\ell=1}n^{2\ell}\Biggl[\sum^{s}_{m=0}b_{m\ell}\Biggr] = 0\;\; \forall \;\; n = \{1,...,s\},\nonumber\\
    {\rm and} \qquad &\sum^{s}_{\ell=1}n^{2(\ell-r)}\Biggl[\sum^{s}_{m=0}b_{m\ell}\,m^{2r}\Biggr] = 1 \nonumber\\
    \qquad &\forall \;\; n = \{1,...,s\} \;\; \mathrm{and} \;\; r = \{1,...,s\}
\end{align}
from the even terms in $\beta$ (i.e. from the cosine terms). The above two set of linear equations can be solved separately to get the coefficients $a$ and $b$ for arbitrary spin $s$ system. We note that our results are consistent with the previous work on this subject~\cite{VANWAGENINGEN1964250,lehrer-ilamed_1964,Curtright:2014cva}.\\

In appendix~\ref{app_spinmatrices_AMO} we provide explicit expressions for spin-$1$, $2$, and $3$ systems.

%~~~~~~~~~~~~~~~~~~~~
\section{Algorithm}
\label{sec:algo_summary}
%~~~~~~~~~~~~~~~~~~~~

\subsection{Algorithm summary}
Equipped with the analytical solution for both the drift and kick evolution along with arbitrary spin matrix exponential, the full split-step Fourier algorithm proceeds as follows: Starting with the field components $\psi_m({\bx},t)$ at time $t$, they are drifted through a time step $\epsilon/2$ according to
\begin{align}\label{eq:vector_step1}
    i\partial_t\psi_n &= \Biggl[\delta_{nn'}\left(-\frac{1}{2\mu}\nabla^2\right) - i\,g_{ij}\,[\hat{S}_i]_{nn'}\,\nabla_j\Biggr]\psi_{n'}\nonumber\\
    \implies \psi^{(1)}_n({\bx}) &= \int_{\bk}\mathcal{F}^{-1}_{\bk,\bx}\;e^{-i\epsilon\,\bk^2/4\mu}[e^{i\epsilon\, g_{ij}k_{j}\hat{S}_{i}/2}]_{nn'}\,\times\nonumber\\
    &\qquad\qquad\qquad\qquad\int_{\bw}\mathcal{F}_{\bk,\bw}\psi_{n'}({\bw},t)\,.
\end{align}
Here the symbol $\mathcal{F}$ represents Fourier transformation: $\int_{\bw}\mathcal{F}_{\bk,\bw}\,h({\bw},t) = \int\mathrm{d}^3w\, e^{i{\bm k}\cdot{\bm w}} h({\bm w},t) = h_{\bk}(t)$. Similarly $\mathcal{F}^{-1}$ represents inverse Fourier transformation: $\int_{\bk}\mathcal{F}^{-1}_{\bk,\bx}\,h_{\bk}(t) = \int\frac{\mathrm{d}^3k}{(2\pi)^3}\, e^{-i{\bm k}\cdot{\bx}} h_{\bk}(t) = h(\bx,t)$.\footnote{In practice, we work with a Cartesian cubic grid with spatial resolution $\Delta x$ in each direction, with a finite volume $V = (N \Delta x)^3 = L^3$. This leads to $\int_{\bk} \rightarrow V^{-1}\sum_{\bk}$ with ${\bk} = 2(\Delta x)^{-1}\sin(\pi{\bm n}/N)$, and $\delta^{(3)}({\bx}-{\by}) \rightarrow V\delta_{\bx,\by}$.}
\\ \\
Then, every component is kicked through a time step $\epsilon$ according to
\begin{align}\label{eq:vector_step2}
    &i\partial_t\psi_{n} = \left(\mu V - \frac{\lambda}{\mu^2} \rho\right)\psi_{n} + \gamma\,f(t)\,{\bm B}\cdot\bm{\hat{S}}_{nn'}\psi_{n'}\nonumber\\
    &\qquad\qquad - \frac{\alpha}{\mu^2}\bm{\mathcal{S}}\cdot\bm{\hat{S}}_{nn'}\psi_{n'} + \frac{\xi}{2s+1}\hat{A}_{nm}\,\psi^{\ast}_m\psi_{m'}\,\hat{A}_{m'n'},\nonumber\\
    &\implies \psi^{(2)}_{m}({\bx}) = \hat{\mathcal{U}}_{n\ell}(\epsilon)\,\hat{\mathcal{U}}'_{\ell n}(\epsilon)\,\psi^{(1)}_{n'}({\bx})\,,
\end{align}
where the operators $\hat{\mathcal{U}}_{mn}(\epsilon)$ and $\hat{\mathcal{U}}'_{mn}(\epsilon)$ are given by~\eqref{eq:kick_solution}, with $\rho$, $\bm{\mathcal{S}}$ and $\rho_{q}$ computed using $\bPsi^{(1)}$. In the case where $V$ and $f{\bm B}$ are self generated potentials (c.f. Eq.~\eqref{eq:gravity_Poissoneqn} for self gravity in the cosmological context, while Eq.~\eqref{eq:dipolarfield_Poissoneqn} for self dipolar field in the condensed matter context), they are easily computed in Fourier space as\footnote{In these computations, we always discard the zero momentum mode for practical purposes.}
\begin{align}
    V \rightarrow \Phi({\bm x}) &= -4\pi \mu G\int_{\bk}\mathcal{F}^{-1}_{\bk,\bx}\;\frac{1}{k^2}\int_{\bw}\mathcal{F}_{\bk,\bw}\rho({\bw})\,,\nonumber\\
    %%%%%%%
    f{\bm B} \rightarrow \nabla a({\bm x}) &= -\gamma\int_{\bk}\mathcal{F}^{-1}_{\bk,\bx}\;\frac{{\bm k}}{k^2}\Biggl[{\bm k}\cdot\int_{\bw}\mathcal{F}_{\bk,\bw}\,\bm{\mathcal{S}}({\bw})\Biggr]\,.
\end{align}
\\ \\
Finally, the fields are again drifted through a time step $\epsilon/2$
\begin{align}\label{eq:vector_step3}
    &\qquad i\partial_t\psi_n = \Biggl[\delta_{nn'}\left(-\frac{1}{2\mu}\nabla^2\right) - i\,g_{ij}\,[\hat{S}_i]_{nn'}\,\nabla_j\Biggr]\psi_{n'}\nonumber\\
    &\implies \psi_{n}({\bx},t+\epsilon) = \int_{\bk}\mathcal{F}^{-1}_{\bk,\bx}\,e^{-i\epsilon\,\bk^2/4\mu}[e^{i\epsilon\, g_{ij}k_{j}\hat{S}_{i}/2}]_{nn'}\,\times\nonumber\\
    &\qquad\qquad\qquad\qquad\qquad\qquad\qquad\int_{\bw}\mathcal{F}_{\bk,\bw}\,\psi^{(2)}_{n'}({\bw})\,.
\end{align}
The half drift steps in the set of operations ensure $\mathcal{O}(\epsilon^2)$ accuracy, while successive computation of the kick ensures reversibility. Since every operation is unitary, the full drift-kick-drift algorithm is symplectic and conserves total particle number. For cases with time-independent external potentials, the total spin is also conserved.  

The relevant matrix exponentials appearing in both the kick and drift steps, for a general spin-$s$ system, are obtained as in Eq.~\eqref{eq:ansatz_exp}, augmented with Eq.~\eqref{eq:a_relations} and Eq.~\eqref{eq:b_relations} to get the expansion coefficients. As examples, explicit expressions for spin-$1$, $2$, and $3$ are given in appendix~\ref{app_spinmatrices_AMO}. Relevant for cosmology/field theory, explicit expressions for spin-$1$ and spin-$2$ matrices are given in appendix~\ref{app_spinmatrices_FT}.

\subsection{Courant–Friedrichs–Lewy condition}\label{sec:cfl.condition}

In order to get a reliable field evolution, it must be ensured that the fastest process occurring in the system is sufficiently resolved. The Courant–Friedrichs–Lewy (CFL) condition takes care of this by choosing sufficiently small time step $\epsilon$. 

In the drift step there are two processes. One is the usual free field evolution (due to the Laplacian term) $\sim e^{i(\epsilon/2\mu)\nabla^2/2}$, and another is the SO term $\sim e^{i(\epsilon/2)gs|\nabla|}$ where we have replaced the spin matrix by $s$ (for the maximum spin multiplicity corresponding to the fastest frequency) and $g$ is the largest entry in the matrix $g_{ij}$. 

On the other hand, the kick evolution involves 5 different pieces. Two of them are due to couplings with external (or self generated long ranged) scalar and vector fields, giving $\sim e^{-i\epsilon\mu V}$ and $\sim e^{-i\epsilon\gamma B f s}$ respectively. The other three are due to short-range self-interactions of the \schr field. These give the factors $\sim e^{i\epsilon(\lambda/\mu^2)\rho}$ and $\sim e^{i\epsilon(\alpha/\mu^2)s\mathcal{S}}$ for density-density and spin-spin interactions respectively, while $\sim e^{\pm i\epsilon\xi\rho_q/(2s+1)}$ for the spin singlet interaction. Here recall that $\rho_q = \sqrt{\rho^2 - |q|^2}$ with $q = \psi^{T}A\psi$, and once again we have replaced the spin matrix in the exponent with the largest multiplicity (eigen-)value $s$. With all these 6 different pieces, the CFL condition reads
\begin{align}\label{eq:CFL.cond}
    \epsilon &= 2\pi\, \delta\,\min\Biggl[\frac{\mu}{3}(\Delta x)^{2} \,,\, \frac{\Delta x}{\sqrt{3}\,s\,{\rm max}[g]} \,,\, |\mu V|^{-1} \,,\,|\gamma s f {\bm B}|^{-1} \nonumber\\
    &\qquad\qquad\qquad |\frac{\lambda}{\mu^2}\rho|^{-1} \,,\, |\frac{\alpha}{\mu^2}s\,\bm{\mathcal{S}}|^{-1}\,,\,|\frac{\xi}{2s+1}\rho_q|^{-1}\Biggr]\,.
\end{align}
Here $\delta \ll 1$ is a tuning parameter that dictates the amount by which the fastest oscillation is sampled. In the above, we have replaced $\nabla^2$ by its value on the discrete lattice $\sum^{3}_{i=1}(\Delta x)^{-2}\times 4\sin^2(n_i\pi/N)$, and set $n=N/2$ along with $\sum \rightarrow 3$ in order to maximize the sum over sine functions. For the demonstration of the fidelity of the algorithm, we shall pick $\delta < 1/8$ and $\Delta x$ small enough, so that ${\rm min}[\hdots]=\mu(\Delta x)^2/3$ throughout the duration of the simulation.

%~~~~~~~~~~~~~~~~~~~~~
\section{Scope and applications}\label{sec:scope}
%~~~~~~~~~~~~~~~~~~~~~

Here we discuss some of the applications of our system~\eqref{eq:nonrel_action} in a variety of different contexts. In order to demonstrate the fidelity of our construction and scope of validity, we present some simulation results on both the cold atom and cosmology fronts.

\subsection{Spinor quantum gases}

In a quantum system, the interplay between the spin degrees of freedom and the spatial degrees of freedom often leads to a variety of intriguing phenomena. Spinor quantum gases represent an ideal platform to study such phenomena and, indeed, they have been at the forefront of cold atom research in the past few decades. Here we will consider two specific examples related to spinor BECs. In the first example, the spinor BEC is subjected to an artificial monopole magnetic field and confined in a shell trapping potential; in the second, we consider an untrapped spinor BEC stabalized by the combination of self attraction and the spin-orbit coupling. In both of these cases, we will present specific ground states of the systems, by performing imaginary time evolution.

\subsubsection{Trapped spinor BECs in an effective magnetic monopole}

%~~~~~~~~~~~~~~~~~~~~~~~~
\begin{figure*}[t!]
    \centering
     \includegraphics[width=0.9\textwidth]{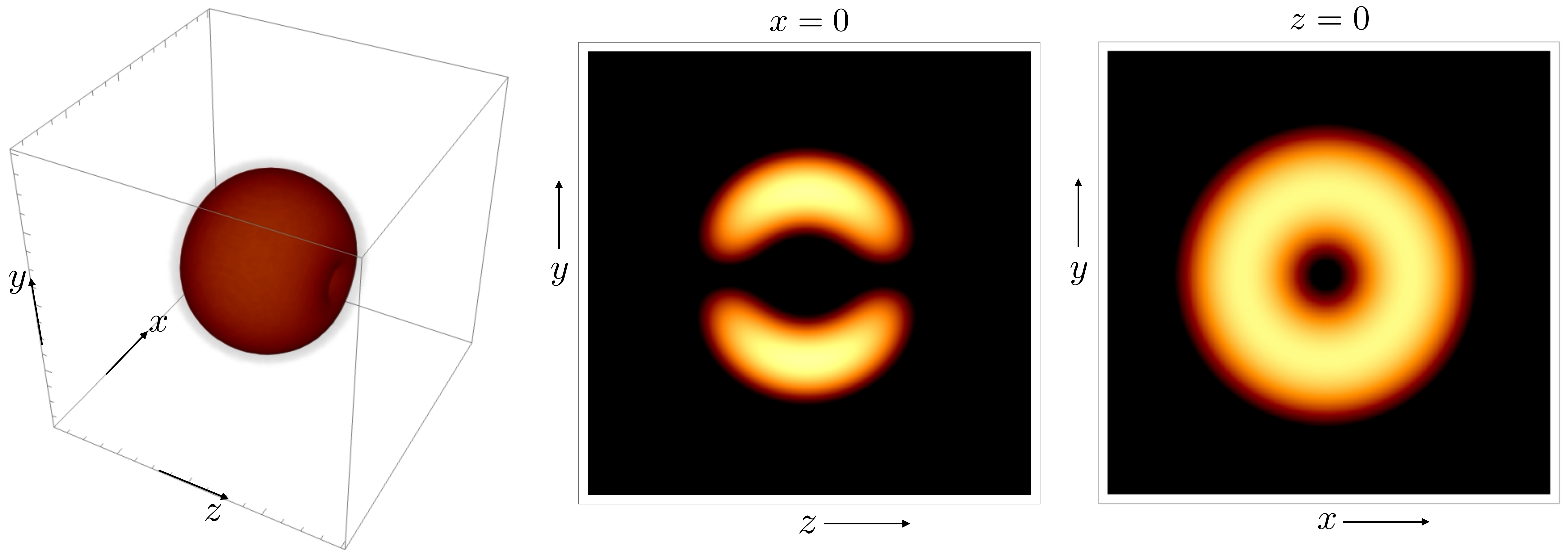}
\caption{A stationary state of the Hamiltonian density~\eqref{eq:Hamiltonian_eff_monopole} with $\lambda = \alpha = 0$, for spin-$1$ system with $m=-1$. Here $\gamma = 4(\mu\omega_T^3)^{1/2}$, the box length is $20$ (in each direction), and the grid size is $61^3$. The left panel gives the full 3D visualization of the number density, whereas the middle and right panels are the number densities as seen in the $y$-$z$ plane (or equivalently $x$-$z$ plane), and $x$-$y$ plane respectively.}
      \label{fig:Monopole_groundstate}
\end{figure*}
%~~~~~~~~~~~~~~~~~~~~~~~~

We first consider the scenario outlined in Ref.~\cite{Zhou:2017xbw}. Atoms with hyperfine spin $s$ are subject to a spherically symmetric harmonic trap, together with a strong bias magnetic field $B_0\hat{\bm z}$ and a periodic quadruple magnetic field $B_1(1-4\tilde{\lambda}\cos\omega t)[x\hat{\bm x} + y\hat{\bm y} - 2z\hat{\bm z}]$. The Zeeman effects due to the bias field can be removed by transforming into the rotating frame along the $z$ axis (rotation frequency being equal to the Larmor frequency $\omega_L$). Then, if $\omega = \omega_L$ and $\tilde{\lambda} = 1$, the magnetic field in the rotating frame has a time-independent piece that is radially outward, mimicking a monopole field, and a fast oscillating piece (with oscillating frequency $\omega = \omega_L$) that can be neglected. The effective
Hamiltonian density in the rotating frame (including self-interactions) turns out to be 
\begin{align}\label{eq:Hamiltonian_eff_monopole}
    \mathcal{H} =& \frac{1}{2\mu} \nabla\psi^{\dagger}_{n}\cdot\nabla\psi_{n} + \mu\,\rho\,V(\bm r) + \gamma\,\bm{\mathcal{S}}\cdot{\bm B}(\bm r)\nonumber\\
    &\qquad - \frac{1}{2\mu^2}\left[\lambda\rho^2 + \alpha\,(\bm{\mathcal{S}}\cdot\bm{\mathcal{S}})\right]\,,
\end{align}
leading to the following spinor \schr equation
\begin{align}\label{eq:eqn_spinorpsi_monopole}
    i\partial_t\psi_n &= \Biggl[\delta_{nn'}\left(-\frac{1}{2\mu}\nabla^2\right) + \left(\mu\,V -  \frac{\lambda}{\mu^2}\,\rho\right)\delta_{nn'}\nonumber\\
    &\qquad + \gamma{\bm B}\cdot\bm{\hat{S}}_{nn'}\;-\; \frac{\alpha}{\mu^2}\,\bm{\mathcal{S}}\cdot\bm{\hat{S}}_{nn'} \Biggr]\psi_{n'}\,.
\end{align}
Here $V({\bm r}) = \omega_T^2\,r^2/2$, ${\bm B} = r\,\hat{\bm r}$, and $\gamma = 2\mu_Bg_FB_1$ (where $\omega_T$ is the harmonic trap frequency, $g_F$ is the Land\'{e} factor, and $\mu_B$ is the Bohr magneton). Owing to the Zeeman coupling of spin density $\bm{\mathcal{S}}$ with ${\bm B}$, we consider configurations where the local spin vectors are polarized opposite to the ${\bm B}$ field. That is, the spinor state is an eigenfunction of the spin operator along the radial direction, and with eigenvalues $m \in [-s,0)$. The full \schr field thus takes the following form
\begin{align}\label{eq:psi_ansatz_eff_monopole}
     \bPsi({\bm r},t) = \phi({\bm r},t)\,\mathcal{M}(\theta,\varphi){\bm \chi}^{(m)}
\end{align}
where ${\bm \chi}^{(m)}$ is the eigenstate of the $\hat{S}_z$ operator with eigenvalue $m$, and $\mathcal{M}(\theta,\varphi) = e^{-i\varphi\hat{S}_z} e^{-i\theta\hat{S}_y}$ is the Unitary transformation matrix that rotates ${\bm \chi}^{(m)}$ to `point' along the radial direction.\footnote{In our $z$ working basis, ${\bm \chi}^{(m)}$ is a column vector with unity at the $m$th position, rest zero. Meaning it is the $m$th column of the matrix $\mathcal{M}$.} With the above ansatz, the effective equation for the scalar field $\phi$ takes the following form\footnote{Here we used the identity $[e^{i\alpha\hat{S}_{y}}]\hat{S}_z[e^{-i\alpha\hat{S}_{y}}] = \hat{S}_z\,\cos\alpha - \hat{S}_x\,\sin\alpha$ (true for all cyclic permutations as well), along with $\sum_i\hat{S}^2_i|{\bm \chi}\rangle = s(s+1)|{\bm \chi}\rangle$ and $\langle{\bm \chi}|\hat{S}_{x,y}|{\bm \chi}\rangle = 0$, to simplify expressions. We also inserted the relationship $\bm{\mathcal{S}} = |\phi|^2\langle{\bm \chi}|\mathcal{M}^{\dagger}\hat{\bm S}\mathcal{M}|{\bm \chi}\rangle = m|\phi|^2\,\hat{\bm r}$. Finally, it is obvious that $\rho = \bPsi^{\dagger}\bPsi = |\phi|^2$.}
\begin{align}\label{eq:phi_eqn_syn_monopole}
    i\partial_t\phi &= - \frac{1}{2\mu}\left(\nabla - im\frac{\cot\theta}{r}\hat{\varphi}\right)^2\phi\nonumber\\
    &\quad + \left(\frac{1}{2}\mu\omega_T^2r^2 + \gamma m r + \frac{\left(s(s+1) - m^2\right)}{2\mu\,r^2}\right)\phi\nonumber\\
    &\quad - \frac{1}{\mu^2}\,\rho\Bigl(\lambda + \alpha m|m|\Bigr)\phi\,.
\end{align}
Ignoring self-interactions for the moment, this dictates the motion of a scalar particle of `electric charge' $m$, in the background of a scalar potential equal to the second term in the first line, and a magnetic monopole at the center (c.f. the vector gauge potential ${\bm A}({\bm r}) = (\cot\theta/r)\,\hat{\varphi}$). To demonstrate our algorithm,  we present the ground state of the above system for the spin-$1$ case and $m =-1$. 
For this purpose, we evolved the Euclidized \schr equation, i.e. $ t \rightarrow -i\tau$ in Eq.~\eqref{eq:eqn_spinorpsi_monopole}, beginning with varying ansatz of the form~\eqref{eq:psi_ansatz_eff_monopole} with $\phi$ being an arbitrary function (typically chosen to be a Gaussian).\footnote{It is to be noted that in the imaginary time evolution, the total `wavefunction' must be re-normalized at every time iteration, for otherwise the total particle number dies out exponentially like $e^{- E_0 \tau}$, where $E_0$ is the ground state energy.} We found convergence towards the ground state shown in~\fref{fig:Monopole_groundstate}. Here we considered $\gamma = 4(\mu\omega_T^3)^{1/2}$. The presence of a cylindrical hole along the $z$ axis is reflective of the gauge potential ${\bm A}$ in the effective system for $\phi$, and can be thought of as a Dirac string. To validate the stationarity and robustness of the obtained ground state, we evolved it in real time and saw no variation (apart from the overall phase rotation). 

If $\gamma \gg (\mu \omega^3_T)^{1/2}(s(s+1)-m^2)^{1/4}(2m^4)^{-1/4}$, the minima of the scalar potential lies at $r_0 \simeq -\gamma m / (\mu\omega_T^2)$ (with $m < 0$), and the spinor field is expected to be tightly concentrated within the spherical shell at this radius. The problem reduces to that of a charged particle confined on a spherical surface subject to a magnetic monopole of charge $m$, centered at the origin~\cite{Zhou:2017xbw}. \\

\noindent{\textit{Inclusion of self-interactions}:} We also investigated the effect of both the spin-dependent and spin-independent self-interactions. The overall effect on the ground state was as expected: When the self-interactions were attractive, the number/spin density compressed, whereas for repulsive self-interactions, the shape of number/spin density ``swelled".

\subsubsection{Self-trapped BECs with Spin-Orbit coupling}

%~~~~~~~~~~~~~~~~~~~~~~~~
\begin{figure*}[t!]
    \centering
     \includegraphics[width=0.7\textwidth]{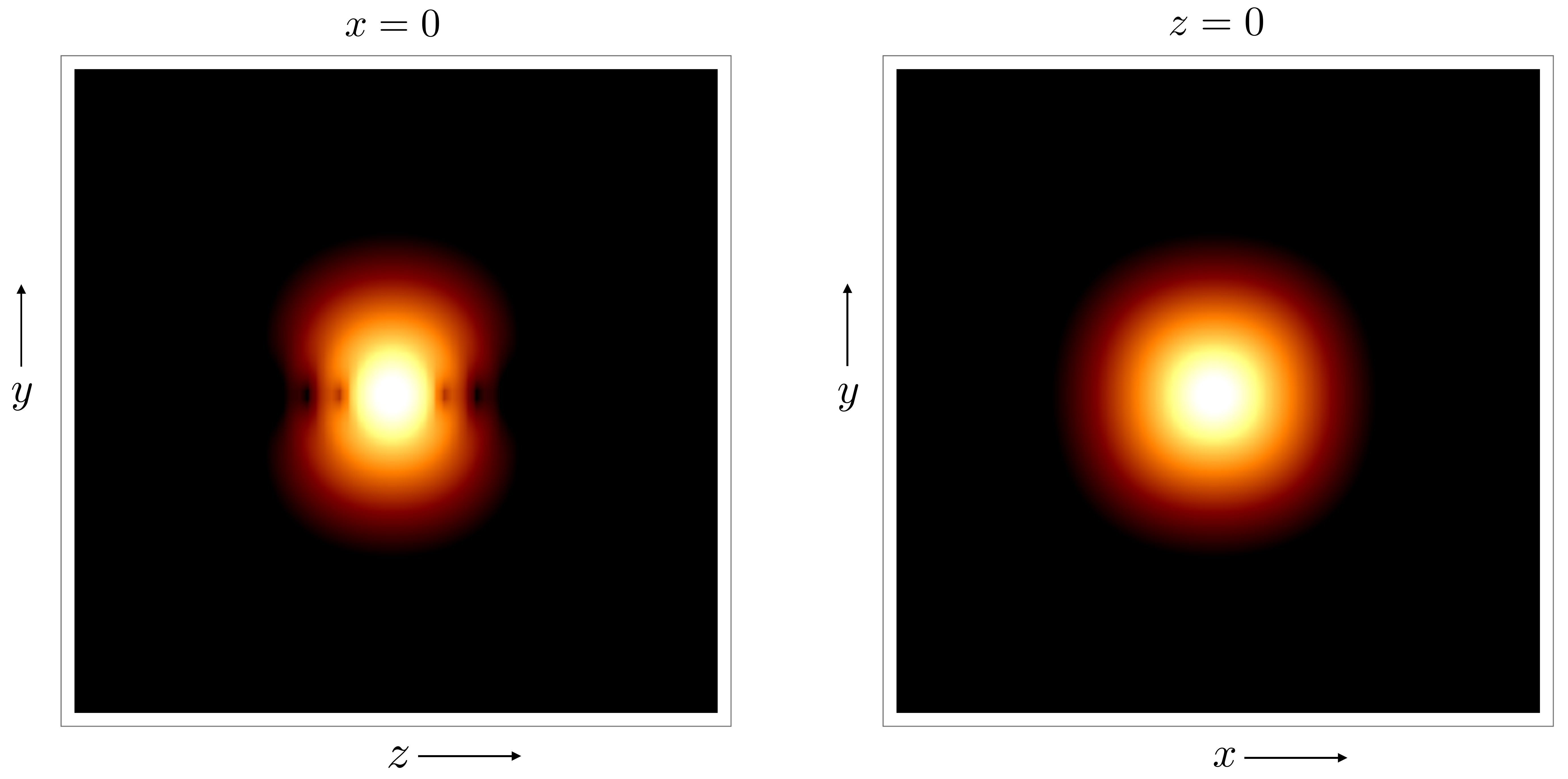}
\caption{A stationary state of the Hamiltonian density~\eqref{eq:Hamiltonian_SOC} for a spin-$1$ system, with $\mu = 1$, $\lambda = 3$, $\alpha = 0$, and $g_{ij} = 2\delta_{ij}$. The box length is $30$ in each direction, and the grid is $81^3$. The state has a cylindrical symmetry (about the $z$ axis). The left panel shows the number density as seen in the $x$-$z$ plane (or equivalently $y$-$z$ plane), whereas the right panel shows the number density in the $x$-$y$ plane.}
      \label{fig:SOC_groundstate}
\end{figure*}
%-----------------------

Our second example concerns the existence of self-trapped spinor BEC with attractive interaction in free space.
Without any confining potentials such as a harmonic trap in the BEC context or gravity in the cosmology context, it is well known that in dimensions 2 and above, the GPE system does not admit bound solitonic states with attractive self-interactions only.\footnote{Although such states can exist in 1 spatial dimension~\cite{PhysRevA.79.013423,2002Natur.417..150S,Nguyen:2014,Luo2020}.} However, a novel way to realize such (quasi-stable) bound states in high dimensions without any trapping potential was presented in Refs.~\cite{PhysRevE.89.032920,PhysRevLett.115.253902} for a two component GPE system, where the stability is provided by means of a spin-orbit (SO) coupling term. 
%(Also see~\cite{PhysRevLett.100.090406} for a similar analysis in 2D.) 
Generalizing the framework to a general spin-$s$ system, the energy is
\begin{align}\label{eq:Hamiltonian_SOC}
    H &= \int\mathrm{d}^3x\,\Biggl[\frac{1}{2\mu} \nabla\psi^{\dagger}_{n}\cdot\nabla\psi_{n} - \frac{1}{2\mu^2}\left(\lambda\rho^2 + \alpha\,(\bm{\mathcal{S}}\cdot\bm{\mathcal{S}})\right) \nonumber\\
    &\qquad\qquad\quad - i\,g_{ij}\,\psi^{\dagger}_{n}\,[\hat{S}_i]_{nn'}\,\nabla_j\,\psi_n\Biggr]\,.
\end{align}
To simplify matters, here we only consider the case when SO coupling operator reduces to the helicity operator, i.e. $g_{ij} = g\delta_{ij}$ giving $g_{ij}\hat{S}_i\nabla_{j} = g\,\hat{\bm S}\cdot\nabla$. In order to analyze the structure of quasi-stable bound states (if any), consider field solutions with some characteristic size $R$ and total particle number $N$. The three different energy terms, corresponding to the usual pressure, self-interactions, and SO coupling become
\begin{align}
    H_{\rm kin} &= c_{\rm kin}\frac{N}{\mu R^2}\,,\nonumber\\
    H_{\rm self} &= -\frac{N^2}{\mu^2R^3}\left(\lambda c_{\rm si} + \alpha c_{\rm sd}\right) \,,\nonumber\\
    H_{\rm so} &= -g c_{\rm so}\frac{N}{R}\,,
\end{align}
with the total energy equal to the sum of the three, $H = H_{\rm kin} + H_{\rm self}+ H_{\rm so}$, and where the different $c$'s are positive constants. It can be easily seen that for a fixed $N$, the energy function (as a function of $R$) admits a local minimum at
\begin{align}
    \mu R = \frac{c_{\rm kin}}{gc_{\rm so}} + \frac{1}{g c_{\rm so}}\left(c_{\rm kin}^2 - 3 g c_{\rm so}(\lambda c_{\rm si} + \alpha c_{\rm sd})N \right)^{1/2}\,,
\end{align}
implying $N < c_{\rm kin}^2(3 g c_{\rm so}(\lambda c_{\rm si} + \alpha c_{\rm sd}))^{-1}$ as the necessary condition for its existence.\footnote{It must be noted that if self-interactions are absent, the assumption of bound states and hence the scaling argument breaks down. This is because in this case the Hamiltonian commutes with the momentum operator, rendering any possible eigenstate of $H$ to be dispersive/non-stationary.}\\

In~\fref{fig:SOC_groundstate} we show a quasi-stable state obtained for the Hamiltonian~\eqref{eq:Hamiltonian_SOC} (by evolving the field with imaginary time starting from a similar initial condition as the previous example, together with re-normalizing the field at every iteration), for a spin-$1$ system. It must be noted that in this case of SO coupling, the corresponding self-source term $\bm{\mathcal{J}}' = i\nabla\times\bm{\mathcal{S}} = \bm{p}\times\bm{\mathcal{S}}$ can lead to a non-conservation of total spin. For instance with reflective boundary conditions, any field packet carrying some spin, reflects off from the boundary with a change in the direction of $\bm{p}$, resulting in a change in $\bm{\mathcal{J}}'$. While we observed a slight non-conservation of total spin in our (real time) SO simulations, we have checked the stability of the quasi-stable state shown in~\fref{fig:SOC_groundstate} by changing the boundary conditions to both periodic and absorptive.\footnote{With total energy $\simeq -0.8$ and absorptive boundaries, the object only lost about  $10^{-4}\%$ of its total norm within $\sim 33$ oscillation cycles (in real time evolution).}

\subsection{Cosmological/Astrophysical systems}\label{sec:cosmology}

In the contemporary universe, dark matter can be described by a classical, non-relativistic, bosonic spin-$s$ field \cite{Jain:2021pnk}. The action in \eqref{eq:nonrel_action} can be used to explore the dynamics of such dark matter. In this section, we briefly explore the applications and limitations of using \eqref{eq:nonrel_action} and our corresponding algorithm for exploring dark matter dynamics in an astrophysical and cosmological context. For simplicity, we consider the case where such dark matter only interacts gravitationally with the rest of the Standard Model, but we will allow for non-gravitational self-interactions within the dark sector itself.\\ 

\noindent{\bf Gravitational Effects:} If the spin-$s$ field determines the dominant energy density in a given region,\footnote{We assume that such a region is small compared to cosmological scales, so cosmological expansion can be ignored. See Sec.~5 in \cite{Jain:2022agt} by some of us on how it can be incorporated in the algorithm.} then the potential $V(\bm{x})\rightarrow \Phi(t,\bm{x})$  can be thought of as the gravitational potential due to the dark matter density itself, which is also the dominant potential determining the dynamics of the dark matter density. Similarly, the $\bar{\bm{B}}(t,\bm{x})$ can be interpreted as the gravitomagnetic field generated by the dark matter field itself. Given our assumption of non-relativistic dark matter, the gravitomagnetic effects are expected to be small. Explicitly, for a spatially localized clump of size $R$ and mass $M$ and with maximal spin $M\hbar/\mu$, the gravitomagnetic term is smaller than the gravitational potential term in Eq.~\eqref{eq:nonrel_action} by a factor of $\lambda_c^2/R^2\ll 1$, where $\lambda_c=\hbar/\mu c$ is the reduced Compton wavelength of the underlying dark matter particle. We caution that additional relativistic corrections beyond the ones included in our action are also present and might be equally or more important -- a more careful analysis is warranted (similar to Ref.~\cite{Salehian:2021khb} in the context of scalars). Furthermore, while there is a spin-orbit coupling term due to relativistic corrections in the gravitational system (see for example, \cite{Porto:2016pyg,Cashen:2016neh}), it is not clear whether the spin-orbit coupling term used in this paper can be directly mapped to that one.

If the dark matter field is a subdominant source of energy in a given region, then potential $V$ can be appended by stronger gravitational potentials due to sources in the vicinity; including for example, a black hole. If also rotating, the $\bar{\bm{B}}$ could then be the gravitomagnetic field of such a source. Such gravitomagnetic effects from a relativistic source can be probed by our system. Nevertheless, care is needed to make sure that we self-consistently include relativistic corrections to the action as the dark matter field probes the associated effects. 
%~~~~~~~~~~~~~~~~~~~~~~~~
\begin{figure*}[t!]
    \centering
     \includegraphics[width=4.5in]{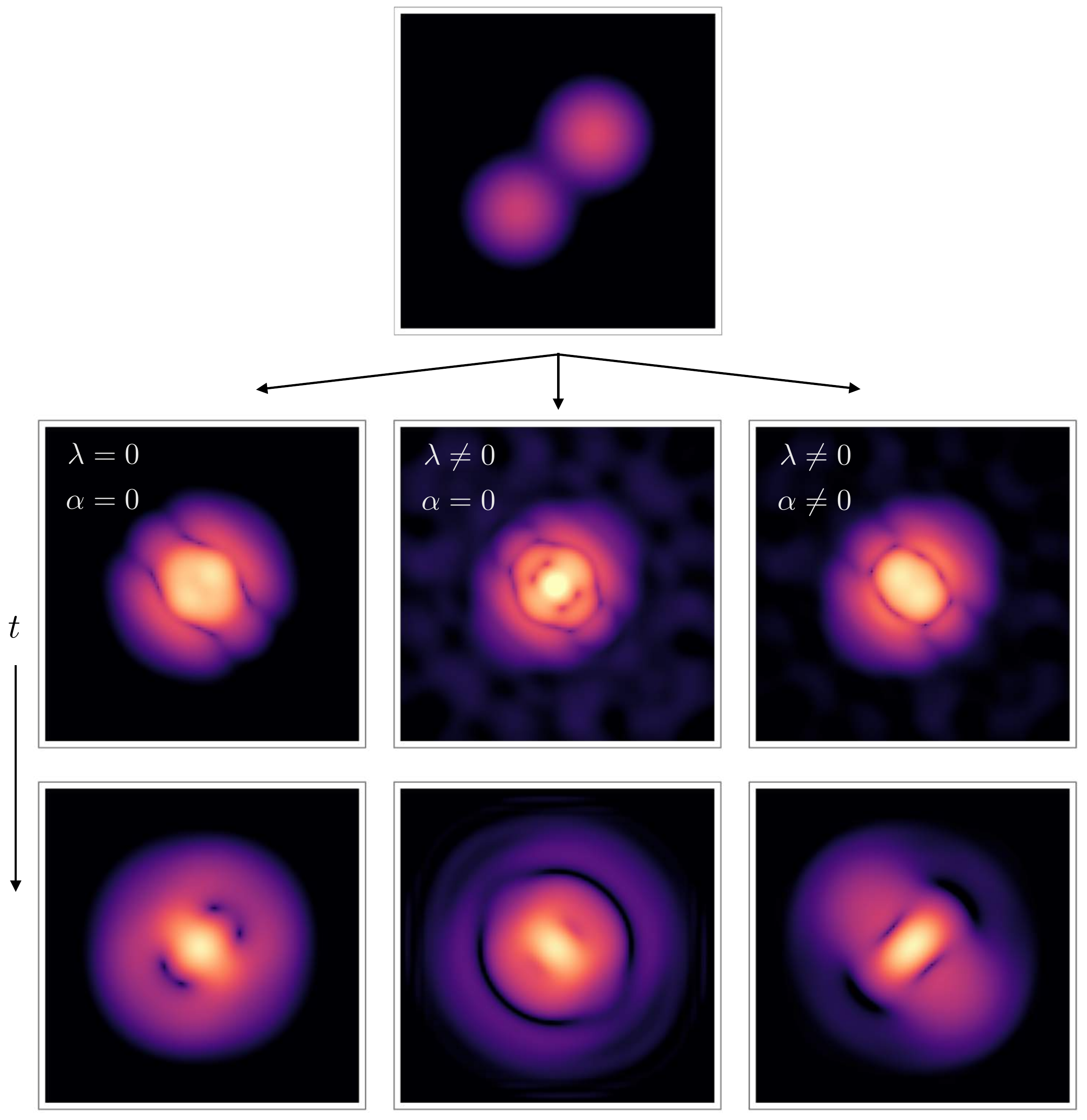}
\caption{Spin density at different instants in the collision of two spin-$1$ solitons, as seen in the $z=0$ plane. With $\mu = 1$ and $G = 1/8\pi$, the box length in each direction is $25$ and the grid size is $101^3$. At the start (top figure), two solitons, each of total `mass' $60$, were only gravitationally bound, were stationary, and diagonally opposite in the $x$-$y$-$z$ space. To capture the effects of gravity, spin-independent, and spin-dependent interactions, we performed three simulations: In left panel, gravitational interactions were included, and point-like interactions were not. Center panel shows the same scenario but with the addition of spin independent (attractive) interaction $\propto \lambda \rho^2$ (with $\lambda = 0.03$). Finally, the right panel shows the case when the spin dependent interaction $\propto \alpha \,\bm{\mathcal{S}}\cdot\bm{\mathcal{S}}$ was also included (with $\alpha = -0.01$). The impact of the spin-independent and spin-dependent interactions are accurately captured by our numerical evolution.}
      \label{fig:solitons_collide}
\end{figure*}
%~~~~~~~~~~~~~~~~~~~~~~~~
\\ \\
\noindent{\bf Including self-interactions:} We now turn to non-gravitational self-interactions of the dark field. The implications of such interactions in an astrophysical/cosmological setting (in particular for higher bosonic fields) have been explored to an extent in earlier papers by some of us \cite{Zhang:2021xxa,Jain:2022kwq,Jain:2022agt}. We review them briefly here, with an eye towards demonstrating the impact of such interactions using our numerical algorithm. For the scalar case with self-interactions and some associated effects, see, for example~\cite{Chavanis:2011zi,Chavanis:2016dab,Amin:2019ums,Chavanis:2020jkc,Dawoodbhoy:2021beb,Shapiro:2021hjp,Chavanis:2022fvh,Mocz:2023adf}.
%To focus on self-interactions, and restricting ourselves to non-relativistic regimes, we will set $V=\Phi$.
 
The precise form of point-like self-interactions in the nonrelativistic limit, is dictated by the UV structure of the bosonic theory. At the quartic level in the IR, both density-density and spin-spin interactions are possible. They are in fact present in some of the usual constructions of interacting spin-$1$ field(s). See for example~\cite{Zhang:2021xxa,Jain:2022kwq} where the quartic interaction term of the vector field $A^{\mu}$ takes the form $\sim (A^{\mu}A_{\mu})^2$, resulting in $\alpha = -\lambda/3$ in the nonrelativistic (IR) limit. Even for the case of a massive spin-$2$ field there are quartic self-interactions~\cite{deRham:2010kj,Hassan:2011hr,Hassan:2011tf,Hassan:2011zd,Hinterbichler:2011tt,deRham:2014zqa,Schmidt-May:2015vnx}, and it could very well be that both density-density and spin-spin interactions are present in the IR.\footnote{The precise value of the $4$-point coupling constant $\lambda$ is dictated by the UV scales. For spin-$1$ case with a Higgs mechanism, $\lambda \sim g^2\mu^2/M^2_{h}$ where $g$ and $M_h$ are the gauge coupling and Higgs mass respectively~\cite{Zhang:2021xxa,Jain:2022kwq}. For spin-$2$ bigravity case, $\lambda \sim \mu^2/\mpl^2$ apart from some overall constants~\cite{Babichev:2016bxi,Jain:2021pnk}.}

These interactions play a significant role in determining the ground state of the system at fixed particle number. In particular, solitons with different spin-multiplicities are degenerate in energy for fixed particle number in absence of self-interactions~\cite{Jain:2021pnk}. However, in the presence of self-interactions the degeneracy gets broken \cite{Zhang:2021xxa,Jain:2022kwq,Jain:2022agt}.

Physics of higher spin solitons, including their emergence time scales and related applications (see for example~\cite{Levkov:2018kau,Eggemeier:2019jsu,Chen:2020cef,Chan:2022bkz} for scalar case and~\cite{Jain:2023ojg,Chen:2023bqy} for spin-$1$ case), merger dynamics and associated production of gravitational waves (see for example~\cite{Helfer:2018vtq} for scalar case while~\cite{Sanchis-Gual:2022mkk} for complex spin-$1$ case), etc. can be strongly affected by point-like self-interactions. There can also arise important differences when considering merger rates of solitons, as well as the eventual configurations of merged objects. Such results are essential for related quantitative astrophysical predictions -- including the small scale mass function in higher spin bosonic dark matter~\cite{Amin:2022pzv}, the generation of electromagnetic radiation from such merged objects~\cite{Amin:2023imi}, etc. %There can also be significant effects of quartic point-like self-interactions on structure formation on small scales. This has only been explored for scalar cases so far, see for example~\cite{Amin:2019ums,Dawoodbhoy:2021beb,Shapiro:2021hjp,Mocz:2023adf}.  

\subsubsection{Numerical examples with gravity and self-interactions}
\noindent{\bf Polarized ground states}: Using our algorithm (with Euclidean time evolution)\footnote{In using Euclidean time evolution to find the ground state, we constantly re-normalize the field at each time step. In the case where non-linearities are present, the value by which one re-normalizes matters. For example, to construct a soliton with total particle number $N$, we re-normalize the field by $\sqrt{N}$ at every iteration.}, we have verified that for a spin-$1$ field with attractive self-interactions as well as gravity, the $0$ spin-multiplicity soliton is the ground state. In the repulsive interaction case, the ground state is the $+1$ (or $-1$) spin multiplicity soliton. This is consistent with our analytical results in Refs.~\cite{Zhang:2021xxa,Jain:2022kwq,Jain:2022agt}.\\

\noindent{\bf Soliton Mergers}: To explore the effect of self-interactions on mergers, we carry out three simulations of binary soliton mergers with identical initial conditions. In all three cases, the initial solitons are identical, are supported by gravitational interactions alone, and with spin density pointing in the $x$ direction. Their centers are located along a diagonal of the $xyz$ co-ordinate system. See top panel of Fig.~\ref{fig:solitons_collide}.

The time evolution of the spin density in the $x$-$y$ plane is shown for the three simulations in the bottom two panels of Fig.~\ref{fig:solitons_collide} (time runs downward). The left most frames include only gravitational interactions. The center frames include gravitational interaction and the spin-independent part of the self-interaction, $\lambda\rho^2$. Lastly, in the right most frames we show results with all the three interactions: gravity, spin-independent interaction $\lambda\rho^2$ and spin-spin interaction $\alpha\,\bm{\mathcal{S}}\cdot\bm{\mathcal{S}}$. We use $\alpha=-\lambda/3$, consistent with the low energy effective theory of the Abelian (heavy-)Higgs model~\cite{Zhang:2021xxa,Jain:2022kwq}. In all three cases, gravity brings the two solitons together. As the profiles overlap, the self-interaction starts playing an important role. The differences in the mergers are evident in the late time frames. It is likely the fraction of mass emitted during the merger, the time-scale of the merger, as well as the final merged object will differ in the three cases (this will be pursued quantitatively elsewhere). We have checked that in all three cases, spin and mass are conserved to machine precision; demonstrating that our algorithm and the corresponding  code deals with self-interactions appropriately.

% %------------------
% \section{Extension towards higher order symplectic integrators}
% \label{Sec:extension}
% %------------------
% \MJ{Here}

% The split-step Fourier method (also known as the partitioned Runge-Kutta method) discussed so far in this paper, is $\mathcal{O}(\epsilon^2)$ accurate where $\epsilon$ is the discrete time step.\footnote{While the error at each step is of the order $\mathcal{O}(\epsilon^3)$, the accumulated error grows and the full evolution of the field is only $\mathcal{O}(\epsilon^2)$ accurate.}. This can be easily extended towards some of the known higher order accurate methods. With the full Hamiltonian broken into a drift and kick hamiltonian piece (c.f. Eq~\eqref{eq:H_drift_n_kick}), a fourth order accurate method, for instance, can be obtained by evolving the field as follows

% over a time step. The $a$ and $b$ coefficients are given in Table $2$ of~\cite{asdf}, with $a_{8-i} = a_i$ and $b_{7-i} = b_i$. Similarly, a sixth order

%------------------
\section{Summary and Discussion}
\label{Sec:summary}
%------------------

In this paper we have devised a symplectic algorithm employing the well-known split-Fourier technique to evolve arbitrary spin-$s$ Gross-Pitaevski systems that are relevant for both AMO systems and astrophysics/cosmology.
The multicomponent/spinor \schr field $\bPsi$ transforms as a vector, in the $2s+1$ dimensional (irreducible) Unitary representation of SO($3$). With analytic closed form expressions for arbitrary spin matrix exponentials, 
we can simulate arbitrary spin-$s$ non-linear \schr systems containing many different types of field interactions of general interest.\\

We consider interactions up to quartic order in the field $\bPsi$, and to leading order in the non-relativistic limit. At the quadratic level, these include interactions of the field with external scalar and vector potentials (both of which can be spatially varying in general). For example in the case of AMO systems, the external potentials include harmonic traps and magnetic fields, while in the case of cosmology they include external gravitational potential generated by some source.

At the quartic level, we include several interactions, both long and short-ranged. For the long-range interactions, we can have the dipolar ($\sim 1/r^3$) self-generated potential in the case of spinor AMO systems, while the Newtonian gravitational ($\sim 1/r$) self- generated potential in the case of ultralight dark matter in cosmology. 

For short-ranged (point-like) interactions, we include both spin-independent and spin-dependent interactions. The former is proportional to the square of the number density, while the latter can be further subdivided into two types: spin-spin interaction being proportional to the norm of the spin density squared, and spin-singlet interaction involving $2$ particle spin-singlet in and out states. Such interactions are of interest in both AMO systems and integer-spin self-interacting dark matter cosmology.

Lastly, we also include the well known spin-orbit coupling term relevant for many AMO systems targeted towards studying spinor BECs.\\

We discussed possible applications of our work in a variety of different contexts, both in the study of spinor BECs in AMO physics and dark matter cosmology. For demonstration purposes, we present some simulation results on both of these fronts. 

For the AMO case, we present ground states for two different scenarios. In the first scenario, the spinor ultracold atomic gas is subject to a synthetic ``hedgehog" magnetic field along with a radially symmetric harmonic trap. This leads to the trapping of the field onto a spherical shell along with the emergence of Dirac strings due to an effective `magnetic monopole' field (where the spin/hyperfine quantum number of the atoms acts as the charge). In the second scenario, the spinor gas has attractive quartic self-interactions and is only subject to a spin-orbit coupling. Such a setup leads to the creation of quasi-stable self-bound solitonic states owing to the balancing of gradient pressure with the attractive self-interactions and spin-orbit coupling induced attraction together. 

For the case of cosmology, we present examples of binary mergers of spin-$1$ solitons, with a focus on the role played by the different interactions: long-range gravitational self-interactions, and short-range spin-independent and spin-dependent self-interactions. We see features in the three collision cases which are reflective of the distinct nature of each of the interactions.\\

The split-step Fourier method (also known as the partitioned Runge-Kutta method) discussed in this paper is $\mathcal{O}(\epsilon^2)$ accurate, where $\epsilon$ is the discrete time step.\footnote{While the error at each step is of the order $\mathcal{O}(\epsilon^3)$, the accumulated error grows and the full evolution of the field is only $\mathcal{O}(\epsilon^2)$ accurate.} With the full Hamiltonian broken into a drift and kick hamiltonian piece (c.f. Eq.~\eqref{eq:H_drift_n_kick}), the accuracy of the integrator can be extended by applying the kick and drift operations in succession, with appropriately chosen coefficients in the respective exponents~\cite{BLANES2002313}. We believe that the method we developed here can be used in a wide range of fields.

\acknowledgments

MJ would like to thank Claudio Castelnovo (Cambridge University) for useful discussions. MAA and MJ are supported by a NASA grant 80NSSC20K0518, HP is supported by NSF PHY-2207283 and the Welch Foundation Grant No. C-1669.

%%%%%%%%%%%%%%%%%%%%%%%%%%%%%%%%%%%%%%%%%%%%%%%%%
%------------------
% References
%------------------
\bibliography{reference}
%%%%%%%%%%%%%%%%%%%%%%%%%%%%%%%%%%%%%%%%%%%%%%%%%
%\newpage
\appendix

%------------------
\section{Conventional spin matrices for bosonic field theories}\label{app_spinmatrices_FT}
%------------------

From a field theoretic point of view, and specializing towards massive vector (spin-$1$) and tensor (spin-$2$) case, the field components are usually expressed in Cartesian basis. The spin angular momentum in the non-relativistic limit is~\cite{Jain:2021pnk}
\begin{align}\label{eq:spin_cartesian}
     \mathcal{S}_{k} = s\,i\,\varepsilon_{ijk}[\bPsi\,\bPsi^{\dagger}]_{ij}\,.
\end{align}
Here $\varepsilon$ is the Levi-Civita symbol, and the quantity $[\bPsi\,\bPsi^{\dagger}]_{ij} = \psi_i\psi^{\ast}_{j}$ for the vector, while $[\bPsi\,\bPsi^{\dagger}]_{ij} = \psi_{ik}\psi^{\ast}_{jk}$ for the tensor case. From this, we can obtain the spin matrices by decomposing the field $\bPsi$ in the spin basis using the polarization vectors/tensors~\cite{Jain:2021pnk}
\begin{align}\label{eq:field_spinbasis}
    \bPsi = \sum^{s}_{m = -s}\psi_{m}\,\bepsilon^{(m)}_{s,\hat{n}}\,.
\end{align}
The set $\{\bepsilon^{m}_{s,\hat{n}}\}$ is orthogonal and complete. That is, we have
\begin{align}\label{eq:orthonormality_condition}
   \Tr[{\bepsilon}^{(m')\;\dagger}_{s,\hat{n}}{\bepsilon}^{(m)}_{s,\hat{n}}] &= \delta_{m'm}\,,\nonumber\\
   \sum_{m}\left[{\bepsilon}^{(m)}_{s,\hat{n}}\;{\bepsilon}^{(m)\dagger}_{s,\hat{n}}\right]_{ij} &= \frac{2s+1}{3}\delta_{ij}\,.
\end{align}
Using the ansatz~\eqref{eq:field_spinbasis} into~\eqref{eq:spin_cartesian}, and identifying $\mathcal{S}_{k} \equiv \psi^{\ast}_{m'}[\hat{S}'_{k}]_{m'm}\psi_m$, we get the following form for the spin matrices
\begin{align}\label{eq:spin_matrices2}
    [\hat{S}'_{k}]_{mm'} = s\,i\,\varepsilon_{ijk}[\bepsilon^{(m')}_{s,\hat{n}}\,\bepsilon^{(m)\,\dagger}_{s,\hat{n}}]_{ij}\,.
\end{align}
Working with the explicit forms of the polarization tensors $\bepsilon$ for spin-$1$ and spin-$2$ case respectively, it can be seen that the spin matrices~\eqref{eq:spin_matrices2} indeed have the desired Lie-algebra of the $SO(3)$ group, and the total spin squared matrix is equal to $s(s+1)$ times the identity. That is,
\begin{align}\label{eq:relations_2}
    [\hat{S}'_x,\hat{S}'_y] &= i\hat{S}'_z\qquad\rm{with\,all\,cyclic\,permutations.}\nonumber\\
    \hat{\bm S}'\cdot\hat{\bm S}' &= s(s+1)\mathbb{I}_{N\times N}\;;\qquad N = 2s+1\,.
\end{align}
In the next two subsections we give the explicit forms for the spin-$1$ and $2$ case.

\subsection{Spin-$1$ case}

In our working ($z$) basis, the polarization vectors $\bepsilon$ take the following conventional form~\cite{Jain:2021pnk} 
\begin{align}
    \bepsilon^{(\pm 1)}_{1,\hat{z}} = \frac{1}{\sqrt{2}}
    \begin{pmatrix}
    1\\
    \pm i\\
    0
    \end{pmatrix}\,;\qquad
    \bepsilon^{(0)}_{1,\hat{z}} =\begin{pmatrix}
    0\\
    0\\
    1
    \end{pmatrix}\,.
\end{align}
Using these in Eq.~\eqref{eq:spin_matrices2}, we get the following explicit forms for the spin matrices
\begin{align}\label{eq:spinmatrices_spin1_FT}
    \hat{S}'_x &= 
    \frac{1}{\sqrt{2}}\begin{pmatrix}
    0 & -1 & 0\\
    -1 & 0 & 1\\
    0 & 1 & 0
    \end{pmatrix}\;,\;
    \hat{S}'_y = 
    \frac{i}{\sqrt{2}}\begin{pmatrix}
    0 & 1 & 0\\
    -1 & 0 & -1\\
    0 & 1 & 0
    \end{pmatrix}\nonumber\\
    \hat{S}'_z &= 
    \begin{pmatrix}
    1 & 0 & 0\\
    0 & 0 & 0\\
    0 & 0 & -1
    \end{pmatrix}\,.
\end{align}
It can be seen that with the above, we do have the relations~\eqref{eq:relations_2} satisfied.

\subsection{Spin-$2$ case}

For the tensor case, the polarization tensors can be obtained using the spin-$1$ polarization vectors as
\begin{align}
    \bepsilon^{(\pm 2)}_{2,\hat{z}} &= \frac{1}{\sqrt{2}}\Bigl(\epsilon^{(\pm 1)}_{1,\hat{z}}\otimes\epsilon^{(\pm 1)}_{1,\hat{z}}\Bigr)\nonumber\\
    %%%%%%%%%%%%%%%%
    \bepsilon^{(0)}_{2,\hat{z}} &= \frac{1}{\sqrt{6}}\Bigl(2\epsilon^{(0)}_{1,\hat{z}}\otimes\epsilon^{(0)}_{1,\hat{z}} - \epsilon^{(1)}_{1,\hat{z}}\otimes\epsilon^{(-1)}_{1,\hat{z}} - \epsilon^{(-1)}_{1,\hat{z}}\otimes\epsilon^{(1)}_{1,\hat{z}}\Bigr)\nonumber\\
    %%%%%%%%%%%%%%%%
    \bepsilon^{(\pm 1)}_{2,\hat{z}} &= \frac{1}{\sqrt{2}}\Bigl(\epsilon^{(0)}_{1,\hat{z}}\otimes\epsilon^{(\pm 1)}_{1,\hat{z}} + \epsilon^{(\pm 1)}_{1,\hat{z}}\otimes\epsilon^{(0)}_{1,\hat{z}}\Bigr)\,,
\end{align}
and take the following form~\cite{Jain:2021pnk}
\begin{align}\label{eq:basis_tensors}
    \bepsilon^{(\pm 2)}_{2,\hat{z}} &= \dfrac{1}{2}\begin{pmatrix}
    1 && \pm i && 0\\
    \pm i && -1 && 0\\
    0 && 0 && 0
    \end{pmatrix}\nonumber\\
    \bepsilon^{(\pm 1)}_{2,\hat{z}} &= \dfrac{1}{2}\begin{pmatrix}
    0 && 0 && 1\\
    0 && 0 && \pm i\\
    1 && \pm i && 0
    \end{pmatrix}\nonumber\\
    \bepsilon^{(0)}_{2,\hat{z}} &= \dfrac{1}{\sqrt{6}}\begin{pmatrix}
    -1 && 0 && 0\\
    0 && -1 && 0\\
    0 && 0 && 2
    \end{pmatrix}\,.
\end{align}
With these, the spin matrices evaluate to be (c.f. Eq.~\eqref{eq:spin_matrices2})
\begin{align}\label{eq:spinmatrices_spin2_FT}
    \hat{S}'_x &= 
    \begin{pmatrix}
    0 & -1 & 0 & 0 & 0\\
    -1 & 0 & -\sqrt{\frac{3}{2}} & 0 & 0\\
    0 & -\sqrt{\frac{3}{2}} & 0 & \sqrt{\frac{3}{2}} & 0\\
    0 & 0 & \sqrt{\frac{3}{2}} & 0 & 1\\
    0 & 0 & 0 & 1 & 0
    \end{pmatrix}\nonumber\\
    \hat{S}'_y &= 
    i\begin{pmatrix}
    0 & 1 & 0 & 0 & 0\\
    -1 & 0 & \sqrt{\frac{3}{2}} & 0 & 0\\
    0 & -\sqrt{\frac{3}{2}} & 0 & -\sqrt{\frac{3}{2}} & 0\\
    0 & 0 & \sqrt{\frac{3}{2}} & 0 & -1\\
    0 & 0 & 0 & 1 & 0
    \end{pmatrix}\nonumber\\
    \hat{S}'_z &= 
    \begin{pmatrix}
    2 & 0 & 0 & 0 & 0\\
    0 & 1 & 0 & 0 & 0\\
    0 & 0 & 0 & 0 & 0\\
    0 & 0 & 0 & -1 & 0\\
    0 & 0 & 0 & 0 & -2
    \end{pmatrix}\,.
\end{align}
Once again it can be easily seen that the relations~\eqref{eq:relations_2} are satisfied.

%------------------
\section{Explicit matrix exponentials for spin-$1$, spin-$2$, and spin-$3$ cases}\label{app_spinmatrices_AMO}
%------------------

\subsection{Spin-$1$ case}

For spin $1$ systems, the exponential matrix has the following analytical solution\footnote{For spin-$1/2$ case where $\hat{\bm S} = \hat{\bm \sigma}/2$, the matrix exponential is the same as~\eqref{eq:exponentiation_s1} with the replacements $\mathbb{I}_{3\times 3} \rightarrow \mathbb{I}_{2\times 2}$, $\hat{\bm n}\cdot\bm{\hat{S}} \rightarrow 2\,\hat{\bm n}\cdot\bm{\hat{S}}$ and $\beta \rightarrow \beta/2$ in the right hand side of~\eqref{eq:exponentiation_s1}.}
\begin{align}\label{eq:exponentiation_s1}
    e^{-i\beta\,\hat{\bm n}\cdot\bm{\hat{S}}} = \mathbb{I}_{3\times 3} - i(\hat{\bm n}\cdot\bm{\hat{S}})\sin\beta + (\hat{\bm n}\cdot\bm{\hat{S}})^2(-1 + \cos\beta)\,,
\end{align}
where
\begin{align}
    \hat{\bm n}\cdot\bm{\hat{S}} = 
    \left(
\begin{array}{ccc}
 n_z & \frac{n_x}{\sqrt{2}}-\frac{i n_y}{\sqrt{2}} & 0 \\
 \frac{n_x}{\sqrt{2}}+\frac{i n_y}{\sqrt{2}} & 0 & \frac{n_x}{\sqrt{2}}-\frac{i n_y}{\sqrt{2}} \\
 0 & \frac{n_x}{\sqrt{2}}+\frac{i n_y}{\sqrt{2}} & -n_z \\
\end{array}
\right)\,.
\end{align}

\subsection{Spin-$2$ case}

For spin-$2$ systems, we have the following closed form expression
\begin{widetext}
\begin{align}\label{eq:exponentiation_s2}
    e^{-i\beta\,\hat{\bm n}\cdot\bm{\hat{S}}} =& \; \mathbb{I}_{5\times 5} + i(\hat{\bm n}\cdot\bm{\hat{S}})\left(-\frac{4}{3}\sin\beta + \frac{1}{6}\sin 2\beta\right) + (\hat{\bm n}\cdot\bm{\hat{S}})^2\left( - \frac{5}{4} + \frac{4}{3}\cos\beta - \frac{1}{12}\cos 2\beta\right)\nonumber\\
    & + i(\hat{\bm n}\cdot\bm{\hat{S}})^{3}\left(\frac{1}{3}\sin\beta - \frac{1}{6}\sin 2\beta\right) + (\hat{\bm n}\cdot\bm{\hat{S}})^4\left(\frac{1}{4} - \frac{1}{3}\cos\beta + \frac{1}{12}\cos 2\beta\right)\,,\nonumber\\
    \nonumber\\
    {\rm where}\qquad
    \hat{\bm n}\cdot\bm{\hat{S}} =& \left(
    \begin{array}{ccccc}
    2 n_z & n_x-i n_y & 0 & 0 & 0 \\
    n_x+i n_y & n_z & \sqrt{\frac{3}{2}} n_x-i \sqrt{\frac{3}{2}} n_y & 0 & 0 \\
    0 & \sqrt{\frac{3}{2}} n_x+i \sqrt{\frac{3}{2}} n_y & 0 & \sqrt{\frac{3}{2}} n_x-i \sqrt{\frac{3}{2}} n_y & 0 \\
    0 & 0 & \sqrt{\frac{3}{2}} n_x+i \sqrt{\frac{3}{2}} n_y & -n_z & n_x-i n_y \\
    0 & 0 & 0 & n_x+i n_y & -2 n_z \\
    \end{array}
\right)\,.
\end{align}
\end{widetext}

\subsection{Spin-$3$ case}

For spin-$3$ case, we get the following closed form expression
\begin{widetext}
    \begin{align}\label{eq:exponentiation_s3}
    e^{-i\beta\,\hat{\bm n}\cdot\bm{\hat{S}}} =& \; \mathbb{I}_{7\times 7} + i(\hat{\bm n}\cdot\bm{\hat{S}})\left(-\frac{3}{2}\sin\beta + \frac{3}{10}\sin 2\beta - \frac{1}{30}\sin 3\beta\right) + (\hat{\bm n}\cdot\bm{\hat{S}})^{2}\left(- \frac{49}{36} + \frac{3}{2}\cos\beta - \frac{3}{20}\cos 2\beta + \frac{1}{90}\cos 3\beta\right)\nonumber\\
    & + i(\hat{\bm n}\cdot\bm{\hat{S}})^{3}\left(\frac{13}{24}\sin\beta - \frac{1}{3}\sin 2\beta + \frac{1}{24}\sin 3\beta\right) + (\hat{\bm n}\cdot\bm{\hat{S}})^{4}\left(\frac{7}{18} - \frac{13}{24}\cos\beta + \frac{1}{6}\cos 2\beta - \frac{1}{72}\cos 3\beta\right)\nonumber\\
    & + i(\hat{\bm n}\cdot\bm{\hat{S}})^{5}\left(-\frac{1}{24}\sin\beta + \frac{1}{30}\sin 2\beta - \frac{1}{120}\sin 3\beta\right) + (\hat{\bm n}\cdot\bm{\hat{S}})^{6}\left( -\frac{1}{36} + \frac{1}{24}\cos\beta - \frac{1}{60}\cos 2\beta + \frac{1}{360}\cos 3\beta\right)\,,
\end{align}
\end{widetext}
where
\begin{widetext}
\begingroup\makeatletter\def\f@size{7}\check@mathfonts
\def\maketag@@@#1{\hbox{\m@th\large\normalfont#1}}%
\begin{align}
    \hat{\bm n}\cdot\bm{\hat{S}} =
    \left(
\begin{array}{ccccccc}
 3 n_z & \sqrt{\frac{3}{2}} n_x-i \sqrt{\frac{3}{2}} n_y & 0 & 0 & 0 & 0 & 0 \\
 \sqrt{\frac{3}{2}} n_x+i \sqrt{\frac{3}{2}} n_y & 2 n_z & \sqrt{\frac{5}{2}} n_x-i \sqrt{\frac{5}{2}} n_y & 0 & 0 & 0 & 0 \\
 0 & \sqrt{\frac{5}{2}} n_x+i \sqrt{\frac{5}{2}} n_y & n_z & \sqrt{3} n_x-i \sqrt{3} n_y & 0 & 0 & 0 \\
 0 & 0 & \sqrt{3} n_x+i \sqrt{3} n_y & 0 & \sqrt{3} n_x-i \sqrt{3} n_y & 0 & 0 \\
 0 & 0 & 0 & \sqrt{3} n_x+i \sqrt{3} n_y & -n_z & \sqrt{\frac{5}{2}} n_x-i \sqrt{\frac{5}{2}} n_y & 0 \\
 0 & 0 & 0 & 0 & \sqrt{\frac{5}{2}} n_x+i \sqrt{\frac{5}{2}} n_y & -2 n_z & \sqrt{\frac{3}{2}} n_x-i \sqrt{\frac{3}{2}} n_y \\
 0 & 0 & 0 & 0 & 0 & \sqrt{\frac{3}{2}} n_x+i \sqrt{\frac{3}{2}} n_y & -3 n_z \\
\end{array}
\right)\,.
\end{align}
\endgroup
\end{widetext}

In general with the relations~\eqref{eq:a_relations} and~\eqref{eq:b_relations} for the coefficients appearing in the exponential of the spin matrices~\eqref{eq:ansatz_exp}, and the expression~\eqref{eq:spin.matrices} for the spin matrices, we can easily get analytical forms for $e^{-i\beta\,\hat{\bm n}\cdot\bm{\hat{S}}}$ for any arbitrary integer spin system. 

\end{document}